\documentclass[openacc]{rsproca_new}

\usepackage{graphicx} 
\usepackage{amsmath,amssymb,latexsym} 
\usepackage{bm} 

\titlehead{Research}

\begin{document}

\title{Programming evolution of geometry in shape-morphing sheets via spatiotemporal activation}

\author{
Daniel Duffy$^{1,\;2}$, Itay Griniasty$^{3}$,
John Biggins$^{1}$ and Cyrus Mostajeran$^{2}$}

\address{$^{1}$Department of Engineering, University of Cambridge,Cambridge CB2 1PZ, UK\\
$^{2}$School of Physical and Mathematical Sciences, Nanyang Technological University, Singapore 637371, Singapore\\
$^{3}$Laboratory of Atomic and Solid State Physics,
Cornell University, Ithaca, New York 14853-2501, USA}

\subject{mechanics, geometry, applied mathematics}


\corres{Cyrus Mostajeran\\
\email{cyrussam.mostajeran@ntu.edu.sg}
}

\begin{abstract}
Shape-programmed sheets morph from one surface into another upon activation by stimuli such as illumination, and have attracted much interest for their potential engineering applications, especially in soft robotics. 
Complex shape changes can be achieved by \textit{patterning} a simple local active deformation (e.g.~isotropic swelling), to generate differential growth. Usually the material itself is designed --- for example by patterning a molecular director --- such that a particular shape change occurs upon exposure to a spatially uniform stimulus. A limitation of this paradigm is that typically only one target geometry can be attained as the stimulus is adjusted. Here we show that this limitation can be overcome by patterning the stimulus itself, thereby exercising spatiotemporal control over local deformation magnitudes. Thus a single physical sample can be induced to traverse a continuous family of target geometries, opening the door to precise shape adjustments, new functionalities, and designable non-reciprocal loops in shape space. We illustrate these possibilities with examples including active parabolic reflectors, chiral flow guides, and bending channels. Finding the necessary patterns of activation involves solving families of metric inverse problems; we solve these by reduction to ODEs in an axisymmetric setting, then present a novel numerical scheme to solve them in generality.
\end{abstract}

\begin{fmtext}
\break\end{fmtext}
\maketitle\noindent

\section{Introduction}

Programmable active solids deform spontaneously in response to an activating stimulus such as heat, light, pressure, or solvent. These materials are particularly exciting when in the form of thin sheets, with patterns of programmed
local in-plane deformation generating complex shape changes upon activation (e.g.~morphing a flat sheet into a conical shell).
This paradigm of `metric mechanics' is common in biology~\cite{thompson1942growth, MahaLongLeaf, forterre2005venus, dervaux2008morphogenesis, alim2016leaf, mitchison2016conformal, serikawa1998analysis, puzey2012evolution}, but can also be realised artificially in gels~\cite{hirokawa1984volume, klein2007shaping, kim2012designing, kim2012thermally, sydney2016biomimetic, na2016grayscale, gladman2016biomimetic, nojoomi2021} or inked polymers \cite{mailen2019thermo} encoded with patterns of isotropic (de)swelling, pneumatic fabrics or elastomers containing patterns of inflatable channels~\cite{martinez2012elastomeric, Pikul2017, siefert2019bio, siefert2020programming, siefert2020inflationary, gao2020shape}, or liquid-crystal elastomers/glasses (LCE/Gs)\cite{warnerbook, warner2020topographic, warner_disclination, aharoni2014geometry, de2012engineering, ware2015voxelated, ambulo2017four, 4dPrintingWare, 4dPrintingSanchez, duffy2021metric, duffy2021shape, barnes2019direct}. These systems hold huge promise for morphing machines and deployable structures, especially given the revolution of 3D printing, and as a result they have become a bustling research area, with potential applications in biomimetics, microfluidics, tissue engineering, and soft robotics.

An important problem that has received little attention is that of the paths through shape space that morphing sheets traverse, e.g.~from a (usually flat) initial configuration to the target surface. Identifying the family of intermediate surfaces that is realised as activation proceeds, and the degree to which these can be designed for, is a fundamental problem of practical significance that is closely linked to functionality in a variety of tasks. For instance, in designing a shape-morphing sheet that activates to take the shape of a parabolic reflector/lens at a given target temperature, the precise shape of the surface at other temperatures is also of practical relevance. Continuing with this particular example, we note that if it is possible to evolve to a paraboloid via a family of paraboloids, the temperature (or other activation parameter) can be used to control focal length. 

In this work we investigate such questions, with a particular focus on systems that offer local control over the orientation of the principal axis of deformation, such as pneumatic~\cite{siefert2019bio, siefert2020programming, siefert2020inflationary} or
nematic LCE/G~\cite{warner2020topographic, warner_disclination, aharoni2014geometry, de2012engineering, ware2015voxelated, ambulo2017four, 4dPrintingWare, 4dPrintingSanchez, duffy2021metric, duffy2021shape} sheets. 
This principal axis, $\hat{\bm{n}}$, is known as the director; we shall take it to be a material property, fixed during fabrication, as is usually the case practically. In contrast, we shall take the magnitude of spontaneous deformation $\lambda$ to be prescribed 
`on the fly', and to depend implicitly on an activating stimulus such as temperature, illumination, or pressure. We do not discuss the specifics of this implicit dependence, which may vary greatly between different experimental implementations, and are essentially downstream of the present work. (We note, however, that the dependence is continuous in pneumatics responding to pressure~\cite{siefert2019bio}, and in nematic LCEs responding to temperature~\cite{tajbakhsh2001,hebner2023discontinuous}, varied via illumination or otherwise.)

\begin{figure}[h]
\centering
\includegraphics[trim={0cm 0cm 0cm 0cm},clip, width=1.0\textwidth]{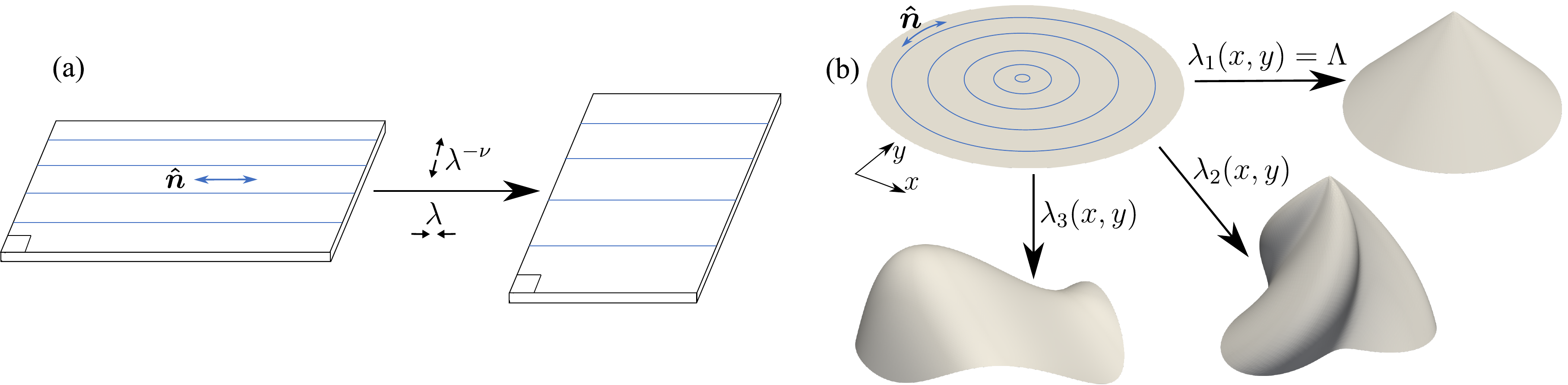}
\caption{(a) A thin sheet with uniform in-plane director $\hat{\bm{n}}$ (blue) spontaneously deforms upon activation. (b) A more complex director (azimuthal here) yields a striking flat $\to$ conical shape change, even for uniform $\lambda = \Lambda$. However, non-uniform $\lambda(x,y)$ patterns enable infinitely many 
other shapes to be generated by the same director pattern.}
\label{fig:intro_sketch}
\end{figure}

Concretely, we will consider thin sheets whose in-plane spontaneous (`programmed') deformation takes the form of a local elongation/contraction by $\lambda$ along $\hat{\bm{n}}$ and $\lambda^{-\nu}$ orthogonally, where the parameter 
$\nu$ is the so-called optothermal Poisson ratio; see Fig.~\ref{fig:intro_sketch}a. The choices $\nu=1/2$, $\nu=0$, and $\nu = -1$ correspond respectively to nematic LCEs, pneumatics, and isotropically swelling gels. 
A line-element vector $\mathrm{d}\bm{l}$ in the unactivated sheet obtains, upon activation, the squared arc length
\begin{equation} \label{eq:nematic metric}
\mathrm{d}l_A^2 = \mathrm{d}\bm{l}\cdot\left(\lambda^2\hat{\bm{n}}\otimes\hat{\bm{n}}+\lambda^{-2\nu}\hat{\bm{n}}_{\perp}\otimes\hat{\bm{n}}_{\perp}\right)\cdot \mathrm{d}\bm{l} \equiv
\mathrm{d}\bm{l}\cdot \bar{a}\cdot \mathrm{d}\bm{l},
\end{equation}
where $\hat{\bm{n}}_{\perp}$ is orthogonal to $\hat{\bm{n}}$ and $\bar{a}$ is the `programmed metric' tensor.
The above is valid in the limit of small sheet thickness, in which case the shape adopted by the activated sheet/surface is a bend-minimising isometry of $\bar{a}$~\cite{efrati2009, lewicka2011scaling, gemmer2013shape, duffy2021metric}.

Typically $\nu$ is a uniform constant while $\hat{\bm{n}}$ is patterned such that interesting shape change occurs as the activating stimulus (and hence $\lambda$) is varied in time but not in space; perhaps  the most canonical example is an LCE disk with purely azimuthal $\hat{\bm{n}}$, which morphs into a cone of semi-angle $\arcsin(\lambda^{1+\nu})$~\cite{ware2015voxelated, modes2011gaussian, ambulo2017four, lopez20184d, guin2018layered}. The general inverse problem of designing $\hat{\bm{n}}$ to encode a given target geometry at a given uniform activation has been successfully tackled computationally~\cite{aharoni2018universal,griniasty2019curved,pauly_2021_inverse}. Then in~\cite{itay_multivalued}, multiple designable features such as director fields and crosslink densities were employed simultaneously to encode multiple target geometries in a single sheet at different values of activating stimulus
--- a `multivalued' inverse problem. While~\cite{itay_multivalued} provided an analytical proof that arbitrary shapes can be achieved via the inverse design of a pressure-like activation, the focus was primarily on programming \textit{material} properties to encode a finite number of target geometries, whereas in the present work we will instead focus on programming \textit{activation} (e.g.~illumination) in a spatiotemporally varying manner, to encode infinitely many geometries in principle; see Fig.~\ref{fig:intro_sketch}b.

While material design/programming relies on technologies that are more mature, the emerging paradigm of activation programming is potentially more powerful, as less is fixed at the time and place of fabrication, leaving more freedom to explore a much richer space of possible shapes and functionalities as the need arises. Despite its experimental complications, activation programming has already been successfully demonstrated in previous works~\cite{kuenstler2020reconfiguring, levin2020self, giudici2022multiple} (the latter also makes significant theoretical contributions). We shall push the relevant theory and numerics forward, mirroring the historical development of director programming by considering high-symmetry cases first, before tackling more general problems.

We will arrive at activation programming by addressing a key motivating question: how can we programme a single morphing sheet to track a \textit{family} of target surfaces (i.e.~a path through shape space)? Examples will demonstrate that material (director) programming is insufficient to achieve this: although a single target affords significant freedom in designing a corresponding director field, this freedom does \textit{not} suffice to encode a continuum of other target geometries in the same sample. Since director patterns usually cannot feasibly be modified on the fly, activation programming is required instead, which is a core message of this work. Mathematically, we must allow $\lambda$ to be a designable function of both space and time, and solve a family of metric inverse problems (one per target geometry). Our main contribution is then a theoretical and numerical framework for solving these inverse problems (i.e.~computing the required spatiotemporal activation) in a general setting.

Roughly speaking, our framework enables us to morph a single sample, fabricated with any fixed director pattern, into essentially any (intrinsic) target geometry/geometries. However, we make an effort throughout to highlight via examples the natural advantages of combining activation programming with material design for a given family of surfaces. For instance, if we seek to attain a prescribed family of surfaces of revolution, it is likely preferable from an implementation viewpoint to achieve this using a circularly symmetric spatiotemporally varying activation field, which becomes feasible provided that the underlying director field possesses the same symmetry. Furthermore, the use of the appropriate director field can facilitate transition to a target surface via a prescribed family using activation fields that become spatially uniform at the target, only exhibiting spatial variation en route.

\subsection*{Paper organization and contributions}

We begin in Sec.~\ref{sec:motivating examples} with several basic examples that illustrate the subtlety associated with the evolution of a shape-morphing sheet to a target surface and the richness that it offers as a space for design in shape programming. We present a general formulation of the metric inverse design problem and provide a brief survey of the literature on the topic in Sec.~\ref{sec: General Metric Inverse Design}. In Sec.~\ref{sec: paraboloid problem} we design a nematic material's director pattern, shape-programming a flat sheet to morph into a \textit{single} paraboloid surface of revolution at a uniform activation $\lambda=\Lambda$. In Sec.~\ref{sec:surfaces of revolution} we then demonstrate how spatiotemporally varying activation $\lambda$ can be used to instead attain a whole \textit{family} of paraboloids.
In Sec.~\ref{sec:general_inverse} we present a novel computational scheme that solves the inverse problem of patterning $\lambda$ to achieve a target geometry in great generality. The power of that computational scheme is then demonstrated through two examples in Sec.~\ref{sec:examples}.

\section{Motivating observations} \label{sec:motivating examples}

We present several simulations illustrating that the evolution of a shape-morphing flat sheet to a target geometry depends strongly on the choice of inverse design (i.e.~the flat domain and the form of $\bar{a}$), and can exhibit \textit{qualitative} transitions. All surfaces are computed using the bespoke software MorphoShell~\cite{MorphoShell_github, duffy2020defective} to minimise a fully nonlinear stretch+bend energy.

\begin{figure}[!h]
\centering\includegraphics[width=\linewidth]{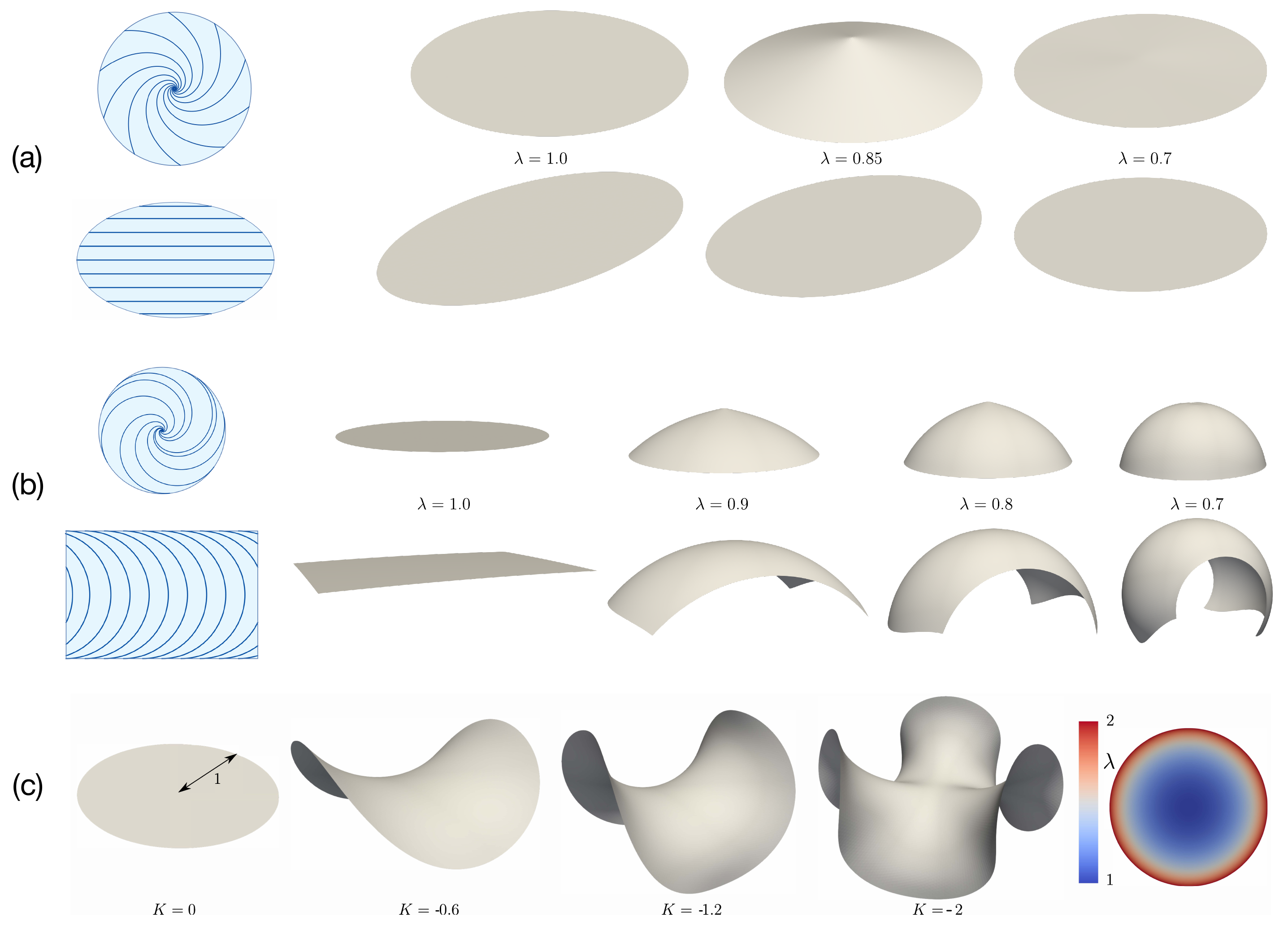}
\caption{(a) Two metrically programmed LCE sheets activate from initially flat states to the same flat target disk, along two qualitatively different paths through shape space. (b) Two metrically programmed LCE sheets activate from initially flat states to different patches of the same sphere. As in (a), the discontinuous +1 topological defect director leads to intermediates containing a sharp tip, while the smooth director yields smooth intermediates. (c) A metrically programmed sheet morphs from a flat disk to a hyperbolic target surface with 3-fold symmetry, exhibiting pronounced qualitative variations between the intermediate surfaces. This evolution resulted from isotropic swelling of the unit disk ($\nu=-1$, $\lambda(r)=4/(4-|K|r^2)$), pictured (right) for $K=-2$.}
\label{figure:combined_1}
\end{figure}

First, consider designing a flat LCE sheet that morphs into a flat target disk at a prescribed uniform activation parameter $\lambda=\Lambda$. Domains and integral curves of two director fields that solve this rather elementary design problem are shown in Fig.~\ref{figure:combined_1}a: a logarithmic spiral (+1 topological defect) inscribed on a disk (different to the target), and a uniform director field (monodomain) inscribed on an ellipse. Intermediate configurations are computed and shown in Fig.~\ref{figure:combined_1}a illustrating that, as $\lambda$ is `dialled' from 1 to $\Lambda$, the uniform-director sample evolves to the target geometry through a family of flat surfaces, whereas the spiral-director sample passes through conical intermediates that rise and fall during the evolution --- a striking and qualitative difference.

Similarly, Fig.~\ref{figure:combined_1}b shows integral curves of two different LCE director fields~\cite{mostajeran2016encoding} that activate to two different patches of the same sphere at $\lambda=\Lambda$. Again the first field is a circularly symmetric +1 topological defect, and is therefore discontinuous, while the second is smooth. The two samples again undergo qualitatively different evolutions, traversing respective families of spindles and spherical patches. Often two designs like these would be considered to solve `the same' metric inverse problem, but they evidently do so in a weaker sense than those in Fig.~\ref{figure:combined_1}a, in that they activate to different subregions of a target. Thus shape-programming inverse solutions come in strong and weak flavours: those that encode a full target, and those that encode only some subregion. Whether the latter type is acceptable or not will be application dependent. 


In Fig.~\ref{figure:combined_1}a and Fig.~\ref{figure:combined_1}b, the Gauss curvature $K$ remains homogeneous throughout activation, apart from at topological defects where sharp tips form (bearing singular $K$) before vanishing at $\lambda = \Lambda$. This intriguing behaviour arises as follows: in regions where $\hat{\bm{n}}$ is smooth, the encoded $K$ is the product of a single $\lambda$-dependent factor and a $\hat{\bm{n}}$-dependent factor~\cite{aharoni2014geometry, mostajeran2015curvature}, while at topological defects an extra $\delta$-function term with a \textit{different} $\lambda$-dependence must be added to the total $K$~\cite{duffy2020defective}, and this topological term vanishes at $\lambda=1$ and $\lambda=\Lambda$. Such singular and non-singular $K$ terms with different $\lambda$-dependencies occur generically where $\hat{\bm{n}}$ is discontinuous, but can occur even for continuous $\hat{\bm{n}}$~\cite{feng2022interfacial}.

If $K < 0$, surprising qualitative transitions can occur during activation even if $K$ is non-singular and homogeneous at all times: In Fig.~\ref{figure:combined_1}c a unit disk (e.g.~of hydrogel~\cite{klein2007shaping}) undergoes inhomogeneous isotropic swelling such that at each time the programmed metric $\bar{a}$ encodes homogeneous $K<0$ (the intrinsic geometry of a hyperbolic disk), with $K$ `dialling' from $0$ to $-2$. It is known that, depending on the value of $K$, the activated surfaces will be either a saddle-shaped patch of a hyperboloid, or one of a sequence of ($n$-fold symmetric) Amsler surfaces~\cite{gemmer2013shape}. Indeed, we observe that as swelling proceeds the shell evolves in a continuous fashion except for sudden transitions between these drastically different forms. The $K$ values at which these transitions occur were calculated in~\cite{gemmer2013shape}.

The above examples demonstrate that the evolution of an activating sheet can be highly non-trivial, and depends on the initial design, with different designs leading to different features that may or may not be desirable. In sections \ref{sec: paraboloid problem}--\ref{sec:surfaces of revolution}, we will study a particular case in more detail, first designing a director pattern to encode a single target geometry, then using spatially inhomogeneous $\lambda$ fields to activate that director pattern and thereby achieve a desired shape evolution that cannot be achieved with homogeneous $\lambda$.

\section{Metric inverse design} \label{sec: General Metric Inverse Design}

Before presenting new results, it will be helpful to formulate the general metric programming problem mathematically, i.e.~the inverse design problem of patterning $\bar{a}$ 
such that a target surface's intrinsic geometry will be attained upon activation. We then review previous work on various versions of this inverse problem, which can be categorised based on the type and number of patternable quantities.

\subsection{Problem formulation}

\begin{figure}[h]
\centering
\includegraphics[trim={0cm 0cm 0cm 0cm},clip, width=1.0\textwidth]{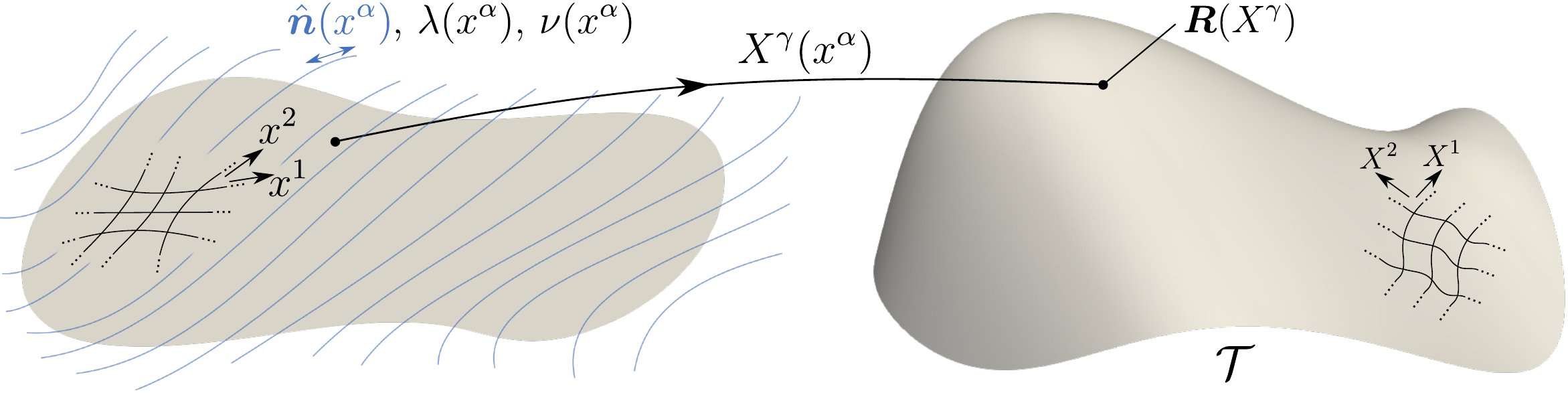}
\caption{Left: A thin unactivated sheet with in-plane director $\hat{\bm{n}}$, contraction factor $\lambda$, and optothermal Poisson ratio $\nu$, all of which could in general vary spatially. Together $\hat{\bm{n}}$, $\lambda$, and $\nu$ parameterise the programmed metric tensor $\bar{a}$ (\ref{eq:nematic metric}). Right: A target surface $\mathcal{T}$, to be attained upon activation. The map $X^\gamma(x^\alpha)$ identifies material points in the unactivated sheet with material points in $\mathcal{T}$. }
\label{fig:general_inverse_sketch}
\end{figure}

For a thin sheet, the general metric inverse design problem can be mathematically formulated as follows: Suppose a target surface $\mathcal{T}$ is inscribed with some coordinates $X^\gamma = (X^1, X^2)$, and has the 
3D position-vector parameterized form $\bm{R}(X^\gamma)$.
Suppose also that the flat plane of the unactivated sheet
is inscribed with coordinates $x^\alpha = (x^1,x^2)$; see Fig.~\ref{fig:general_inverse_sketch}. 
Then any flat $\to$ target mapping $X^\gamma(x^\alpha)$ allows us to pull back the metric tensor of $\mathcal{T}$ to the flat plane, where we denote it by $a$. Writing $\tilde{\bm{R}}(x^\alpha) \equiv \bm{R}(X^\gamma(x^\alpha))$, the components of $a$ with respect to the $x^\alpha$ coordinate basis are
\begin{equation}
a_{\alpha \beta} = \frac{\partial \tilde{\bm{R}}}{\partial x^\alpha} \cdot \frac{\partial \tilde{\bm{R}}}{\partial x^\beta}.
\label{eq:metric_of_general_target}
\end{equation}
Any line element $\mathrm{d}x^\alpha$ in the flat plane corresponds (via the map $X^\gamma(x^\alpha)$) to a line element on $\mathcal{T}$ with squared length
\begin{equation} \label{eq:target_squared_length}
    \mathrm{d}l_{\mathcal{T}}^2 =   \frac{\partial \bm{R}}{\partial X^\gamma} \cdot \frac{\partial \bm{R}}{\partial X^\sigma} \, \mathrm{d}X^{\gamma}
    \mathrm{d}X^{\sigma} =
     a_{\alpha\beta} \, \mathrm{d}x^{\alpha} \mathrm{d}x^{\beta},
\end{equation}
using Einstein summation.

The above quantity can be compared with the 'programmable' squared length $ \bar{a}_{\alpha \beta}  \, \mathrm{d}x^\alpha \mathrm{d}x^\beta$ attained by $\mathrm{d}x^\alpha$ upon activation; see (\ref{eq:nematic metric}).
Thus, if we can find a map $X^\gamma(x^\alpha)$ and an experimentally accessible $\bar{a}_{\alpha \beta}$ such that $a_{\alpha \beta} = \bar{a}_{\alpha \beta}$ throughout some domain, then the inverse problem is solved in that domain, because the target surface's intrinsic geometry will indeed be obtained upon activation (in the small-thickness limit). In general, for solutions to exist, $\bar{a}$ must have at least one tunable degree of freedom (DOF) at each material point, which must be solved for, usually along with the map $X^\gamma(x^\alpha)$.

\color{black}

\subsection{A brief survey of metric inverse problems}
\label{subsec:survey}

The easiest metric inverse problems are those in which the programmable metric $\bar{a}$ has multiple 
DOFs that can be used to `match' the metric of a target surface.
The metric components $a_{\alpha \beta}$ associated with the flat $\to$ target map $X^\gamma(x^\alpha)$ constitute a $2 \times 2$ symmetric matrix field, i.e.~three numbers at each material point. Thus if one can tune three DOFs in $\bar{a}$ at each point (e.g.~parameterized as $\hat{\bm{n}}$, $\lambda$, $\nu$), then one can choose essentially \textit{any} flat $\to$ target map (subject to material feasibility constraints) and set $\bar{a}$ accordingly, making the problem trivial in a sense~\cite{mahaElasticBilayers}. If only two DOFs are available for patterning (e.g.~$\hat{\bm{n}}$, $\lambda$)~\cite{gao2020shape, siefert2019bio}, the problem is harder and more constrained, but still contains significant (indeed surplus) freedom. 

In fact, only one patternable DOF in $\bar{a}$ is required to encode the metric of a target surface, representing the most constrained class of metric inverse problems that are generically solvable. For example, the director $\hat{\bm{n}}$ can be patterned/designed, with the active stretches $\lambda$ and $\lambda^{-\nu}$ being fixed and (usually) homogeneous; in this case the single DOF can be taken to be the angle $\psi$ between $\hat{\bm{n}}$ and the $x$-axis.
High-symmetry targets can be tackled by merely solving ODEs~\cite{aharoni2014geometry, mostajeran2016encoding, warner2018nematic}, as we will do in Sec.~\ref{sec: paraboloid problem}, but otherwise rather involved numerics are employed~\cite{aharoni2018universal, griniasty2019curved, pauly_2021_inverse, aharoni2023plce}. Certain LCE chemistries provide an alternative and highly successful experimental approach, solving the inverse problem mechanically by allowing the material itself to adjust when pressed onto the target shape~\cite{barnes2019direct}. A patterned one-parameter family of shears offers another interesting single-DOF metric programming paradigm, beautifully utilized by \textit{Euglena} algae, and for which ODE-based inverse solutions have been found~\cite{arroyo2014shape, cicconofri2020morphable}.

The most extensively studied single-DOF case is that of isotropic swelling i.e.~$\lambda$ is patterned while $\nu=-1$. This problem is essentially equivalent to the classical one of finding  `isothermal' coordinates that put a target surface's metric into conformal form~\cite{docarmo_curves_surfaces, kreyszig1968introduction}, so naturally it is deeply linked to complex analysis, conformal maps, and harmonic functions.
Furthermore, such parameterizations are extremely useful in computer graphics, so unsurprisingly the general problem has been attacked numerically with much vigor and success~\cite{crane_bff, campen2021efficient, springborn2008conformal, ben2008conformal, mullen2008spectral, zayer2007linear, desbrun2002intrinsic, levy2023least, konakovic2016beyond}. The active solids community has then applied the resulting methods~\cite{nojoomi2021, duque2019distortion}, developed their own~\cite{jones2015optimal}, or instead presented ODE-based solutions~\cite{klein2007shaping, dias2011programmed, warner2019geometry}.

Yet another single-DOF inverse problem is 
that of patterning $\lambda$ to encode the metric of a target surface, given a fixed director field $\hat{\bm{n}}$ and a fixed $\nu$.
Few previous works have discussed this `$\lambda$-inverse' problem, which is a highly nontrivial anisotropic generalization of the isotropic/conformal problem, to which it reduces when $\nu=-1$. As for the other problems discussed above, solutions can be found in high-symmetry cases by merely solving ODEs, as in examples that appear in refs~\cite{cicconofri2020morphable, kuenstler2020blueprinting, giudici2022multiple}, and in our Sec.~\ref{sec:surfaces of revolution}.
With much greater generality, careful application of the method in ref.~\cite{itay_multivalued} can solve the problem (in a PDE initial-value formulation) as long as $\hat{\bm{n}}$ and the target surface are sufficiently smooth, and $\nu \geq 0$. In Sec.~\ref{sec:general_inverse} we will push beyond any such restrictions, presenting a finite-element scheme that solves the fully general case for the first time.

\section{The Paraboloid problem: inverse material design} \label{sec: paraboloid problem}

Due to their unique focusing properties, parabolic reflectors/lenses are widely used in systems that transmit or receive signals: antennas, automobile headlights, solar furnaces, radar and radio relay stations, microphones \ldots etc. Thus inspired, we now turn to the material-design problem of programming a shape-morphing flat sheet to take the shape of a single paraboloid surface of revolution upon spatially uniform activation $\lambda = \Lambda$. Specifically, we will take $\hat{\bm{n}}$ to be the sole patternable DOF in $\bar{a}$.

\subsection{Inverse material design for a target surface of revolution} \label{subsec:inverse_design_rev}

We begin by reviewing the relationship between a target surface of revolution and the required programmable metric in a circularly symmetric design for a nematic LCE or baromorph. The solution to this inverse design problem was originally presented in~\cite{warner2018nematic} in a different form. Here we give a simplified and more direct derivation of the quantities needed to determine the director field $\hat{\bm{n}}$ in the inverse design problem.

We consider a surface of revolution $\mathcal{T}$ given in 3D position-vector parameterized form by $\bm{R}(\rho,\phi)=(\rho\cos\phi,\rho\sin\phi,z(\rho))$, where $\rho,\phi,z$ are cylindrical polar coordinates and the function $z=z(\rho)$ characterizes the surface geometry. From (\ref{eq:target_squared_length}), the squared arc length on the target surface is given by 
\begin{equation} \label{eq:target_arc_rev}
    \mathrm{d}l_{\mathcal{T}}^2=\left(1+z'(\rho)^2\right) \mathrm{d}\rho^2+\rho^2\mathrm{d}\phi^2 ,
\end{equation}
with $(X^1,X^2)=(\rho,\phi)$ parameterizing the surface. Using plane polar coordinates $(r,\theta)$ for the sample's flat unactivated state, we seek to programme the target surface's metric using a circularly symmetric director field described by the angle $\alpha=\alpha(r)$ between $\hat{\bm{n}}$ and the radial direction $\hat{\bm{e}}_r$.

\textbf{Problem I (Inverse design of axisymmetric $\hat{\bm{n}}$):}
\emph{Consider a flat sheet of material that will deform upon activation according to \eqref{eq:nematic metric}, with prescribed uniform magnitudes \(\nu\) and \(\lambda\). Consider also a target surface of revolution defined by $z=z(\rho)$. Find a circularly symmetric director pattern $\alpha(r)$ on the flat sheet such that there exist maps $\rho(r,\theta)$ and $\phi(r,\theta)$  that yield $a_{\alpha\beta}=\bar{a}_{\alpha\beta}$ throughout some region of the flat sheet.}

Using (\ref{eq:nematic metric}) (and as shown in \cite{mostajeran2016encoding}), the programmed metric components are
\begin{align} \label{polarMetric}
\begin{cases}
\bar{a}_{rr}&=\lambda^2+\left(\lambda^{-2\nu}-\lambda^2\right)\sin^2(\alpha),  \\
\bar{a}_{r\theta}&=\bar{a}_{\theta r}=-\frac{1}{2}r\left(\lambda^{-2\nu}-\lambda^2\right)\sin(2\alpha),  \\
\bar{a}_{\theta\theta}&=r^2\left[\lambda^{-2\nu}-\left(\lambda^{-2\nu}-\lambda^2\right)\sin^2(\alpha)\right],
\end{cases}
\end{align}
where the activated squared arc length is $\mathrm{d}l_A^2 = \bar{a}_{rr} \mathrm{d}r^2 + 2\bar{a}_{r\theta} \mathrm{d}r \mathrm{d}\theta + \bar{a}_{\theta\theta} \mathrm{d}\theta^2$. 
Given the circular symmetry of the programmed metric, we consider only axisymmetric deformations of the flat sheet to the target surface of revolution, whereby material circles remain circles and are stretched/compressed uniformly along their length;
i.e.~we consider mappings of the form $\rho = \rho(r)$ and $\phi = \theta + \varphi(r)$. Note that the $\varphi(r)$ term allows for relative rotation of material circles during deformation, a phenomenon that has been noted in theory~\cite{mostajeran2016encoding,warner2018nematic} and observed experimentally~\cite{siefert2020inflationary}.

Switching coordinates to $(r,\theta)$ in (\ref{eq:target_arc_rev}), we obtain
\begin{align} \label{eq:target_arc_rev2}
    \mathrm{d}l_{\mathcal{T}}^2&=\left(1+z'(\rho(r))^2\right) \rho'(r)^2\mathrm{d}r^2+\rho^2(r)\left(\frac{\partial\phi}{\partial r}\mathrm{d}r+\frac{\partial\phi}{\partial\theta}\mathrm{d}\theta\right)^2 \nonumber \\
&=\left[\left(1+z'(\rho(r))^2\right)\rho'(r)^2+\rho(r)^2\left(\frac{\partial\phi}{\partial r}\right)^2\right]\mathrm{d}r^2 + 2\rho(r)^2\frac{\partial\phi}{\partial r}\frac{\partial\phi}{\partial\theta}\mathrm{d}r\mathrm{d}\theta + \rho(r)^2\left(\frac{\partial\phi}{\partial\theta}\right)^2\mathrm{d}\theta^2 \nonumber \\
&=\underbrace{\left[\left(1+z'(\rho(r))^2\right)\rho'(r)^2+\rho(r)^2\varphi'(r)^2\right]}_{a_{rr}(r)}\mathrm{d}r^2 + 2\underbrace{\rho(r)^2\varphi'(r)}_{a_{r\theta}(r)\,=\,a_{\theta r}(r)}\mathrm{d}r\mathrm{d}\theta + \underbrace{\rho(r)^2}_{a_{\theta\theta}(r)}\mathrm{d}\theta^2 
\end{align}
Setting $a_{\theta\theta} = \bar{a}_{\theta\theta}$ and $a_{r\theta} = \bar{a}_{r\theta}$ gives 
\begin{equation}  \label{rho_varphi}
    \rho(r)=\sqrt{\bar{a}_{\theta\theta}(r)} \quad \quad \quad \mathrm{and} \quad \quad \quad \varphi'(r)=\frac{\bar{a}_{r\theta}(r)}{\bar{a}_{\theta\theta}(r)}.
\end{equation}
Substitution into the expression for $\bar{a}_{rr}$ yields
\begin{equation*}
\left(1+z'(\sqrt{\bar{a}_{\theta\theta}})^2\right)\left(\frac{\mathrm{d}}{\mathrm{d}r}\sqrt{\bar{a}_{\theta\theta}}\right)^2+\bar{a}_{\theta\theta}\left(\frac{\bar{a}_{r\theta}}{\bar{a}_{\theta\theta}}\right)^2=\bar{a}_{rr}, 
\end{equation*}
which simplifies to
 \begin{equation} \label{surface of rev solution}
    \frac{\mathrm{d}}{\mathrm{d}r}\bar{a}_{\theta\theta}=\frac{2r\lambda^{1-\nu}}{\sqrt{1+\left(z'\left(\sqrt{\bar{a}_{\theta\theta}(r)}\right)\right)^2}}.
\end{equation}
Eq.~(\ref{surface of rev solution}) can now be used to determine $\alpha(r)$ and thus solve Problem I.

\subsection{The paraboloid case}

Specialising to the case at hand, consider the paraboloid given by $z(\rho) = \frac{1}{2}a\rho^2$,  which has its focus at $(\rho,z)=(0,\frac{1}{2a})$. Eq.~(\ref{surface of rev solution}) at $\lambda=\Lambda$ can be rearranged to give
\begin{equation}
    \Lambda^{1-\nu}r^2 = \int_0^{\bar{a}_{\theta\theta(r)}}\sqrt{1+a^2\bar{a}_{\theta\theta}}\, \mathrm{d}\bar{a}_{\theta\theta} 
    = \frac{2}{3a^2}(1+a^2\bar{a}_{\theta\theta})^{3/2}-\frac{2}{3a^2},
\end{equation}
where we have used $\bar{a}_{\theta\theta}(0)=0$ as a boundary condition for Eq.~(\ref{surface of rev solution}).
Substituting for $\bar{a}_{\theta\theta}$ from (\ref{polarMetric}) and rearranging yields the solution for the director pattern $\alpha(r)$:
\begin{equation} \label{paraboloid director 1}
    \sin^2\alpha(r)
    =\frac{1}{1-\Lambda^{2(1+\nu)}}\left[1-\frac{\Lambda^{2\nu}}{a^2r^2}\left(\left(1+\frac{3}{2}\Lambda^{1-\nu}a^2r^2\right)^{2/3}-1\right)\right] ,
\end{equation}
which is visualised in Fig.~\ref{fig:parabolic_path}a. Note that this solution is only defined out to a finite $r=r_{\max}$ where $\alpha = \pi/2$, limiting the portion/patch of the target paraboloidal profile that can be covered. 

Typically, the programmed sheet will be activated with a spatially uniform $\lambda$ that `dials' from $1$ to $\Lambda$ as, say, illumination is increased. Although the shape achieved at $\lambda=\Lambda$ is indeed the paraboloid we designed for, the intermediate surfaces of revolution are \textit{not} paraboloids.
In fact, they are not even smooth, since the topological defect in the director pattern generates a sharp angular tip for \textit{any} uniform $\lambda$ that is not $1$ or $\Lambda$~\cite{mostajeran2016encoding,kowalski2018curvature, duffy2020defective}. (The spindle in Fig.~\ref{figure:combined_1}b demonstrates exactly the same phenomenon.) Thus the intermediates unfortunately do not offer the useful focusing properties that paraboloids do; this is the problem that we shall remedy in the next section by patterning $\lambda$.

We might suspect that one solution to the above problem could emerge by using a continuous director pattern instead (hence containing no topological defects). However, \textit{any} axisymmetric director field necessarily contains a $+1$ topological defect. Thus, although continuous director designs do exist for a given target paraboloidal geometry and uniform $\lambda = \Lambda$ (and can be found via the PDE-based method of ref.~\cite{griniasty2019curved}; see Supplementary Materials Sec.~S1a), they cannot be axisymmetric. Thus, rather unsurprisingly, the resulting intermediates are not even surfaces of revolution, let alone paraboloids (though they do not exhibit sharp tips). Moreover they cannot retain a circular boundary throughout uniform activation, unlike our axisymmetric design. (On the other hand, due to their lack of symmetry, they do offer a richer design space, arising from the freedom to specify the director along some chosen curves on the target surface, as `initialisation' for the PDE solver.)

\section{Solving the paraboloid problem: tracking a family of surfaces of revolution} \label{sec:surfaces of revolution}

Having solved the uniform-$\lambda$ material-design problem of Sec.~\ref{sec: paraboloid problem}, another quickly presents itself: As mentioned earlier, the director patterns of Sec.~\ref{sec: paraboloid problem} do \textit{not} yield paraboloids when activated with uniform $\lambda$, except at $\lambda = \Lambda$. Can we instead apply a family of spatially varying $\lambda$ fields to attain a family of activated paraboloids? We shall now do so, for example allowing a single LCE sample to be morphed into a whole family of parabolic reflectors with focal length selected via an illumination pattern. In the language of Sec.~\ref{subsec:survey}, we solve a family of $\lambda$-inverse problems, in which $\lambda$ is the sole patternable DOF in $\bar{a}$. For simplicity and tractability, we shall retain circular symmetry, principally utilising the director pattern (\ref{paraboloid director 1}). Similar analysis could be applied for other director patterns, or indeed for families of axisymmetric targets that are not paraboloids.

\subsection{Evolution from a flat sheet using spatiotemporally programmed activation}

Consider a director field $\alpha=\alpha(r)$, and a family of target surfaces of revolution $\mathcal{T}_{\tau}$ parameterized by $\tau$, given by $z_{\tau}=z_{\tau}(\rho_{\tau})$.
The spatiotemporal nature of the activation $\lambda$ 
is accounted for by allowing the metric components in (\ref{polarMetric}) to depend on $(r,\tau)$ through $\lambda=\lambda(r,\tau)$, where 
the parameter $\tau$ (a proxy for time) labels each surface along the evolution to the target. In contrast, the programmed director $\alpha(r)$ is fixed.

\textbf{Problem II (Tracking a family via axisymmetric inverse design of \(\lambda\)):}
\emph{Let there be a flat sheet of material that will deform upon activation according to \eqref{eq:nematic metric}, with prescribed uniform \(\nu\) and a prescribed circularly symmetric director pattern $\alpha(r)$. Consider a family of target surfaces of revolution defined by $z_{\tau}=z_{\tau}(\rho_{\tau})$.
Find circularly symmetric spatiotemporally varying fields \(\lambda(r,\tau)\) such that there exist maps $\rho_{\tau}(r,\theta)$ and $\phi_{\tau}(r,\theta)$ that yield $a_{\alpha\beta}(r, \theta, \tau)=\bar{a}_{\alpha\beta}(r, \theta, \tau)$ throughout some region of the flat sheet.} 

As in Sec.~\ref{sec: paraboloid problem}, we 
consider only axisymmetric deformations, i.e.~mappings of the form $\rho = \rho(r)$ and $\phi = \theta + \varphi(r)$.
Repeating the steps of the derivation in Sec.~\ref{sec: paraboloid problem}, we arrive at
\begin{equation}
    \frac{\mathrm{d}}{\mathrm{d}r}\bar{a}_{\theta\theta}(r,\tau)=\frac{2r\lambda(r,\tau)^{1-\nu}}{\sqrt{1+\left(z_{\tau}'\left(\sqrt{\bar{a}_{\theta\theta}(r,\tau)}\right)\right)^2}},
\end{equation}
which upon substitution from (\ref{polarMetric}) with $\lambda=\lambda(r,\tau)$ yields the equation
\begin{align} \label{DE flat to curved rev}
&\frac{\mathrm{d}\lambda(r,\tau)}{\mathrm{d}r}=
\frac{\lambda}{r\left(\nu c^2- \lambda^{2(1+\nu)}s^2\right)}\times \nonumber \\
& \left[\frac{-\lambda^{1+\nu}}{{\sqrt{1+\left(z_{\tau}'\left(\sqrt{r^2(\lambda^2 s^2
   +\lambda^{-2\nu}c^2)}\right)\right)^2}}}+ 
    r \left(\lambda^{2(1+\nu)}-1\right)\alpha'
   cs + 
    2 \left(\lambda^{2(1+\nu)}s^2 +c^2\right)\right],
   \end{align}
   where $s=\sin\alpha$ and $c=\cos\alpha$.

To find $\lambda(r,\tau)$ for a family of paraboloids given by $z_{\tau}=\frac{1}{2}\tau \rho_{\tau}^2$,
we simply substitute this $z_{\tau}$ and the director field $\alpha(r)$ from (\ref{paraboloid director 1}) into (\ref{DE flat to curved rev}), and solve the resulting differential equation numerically. Since that equation is a first-order ODE, for each $\tau$, we specify $\lambda$ at some $r$ to pick out a unique solution. Even for a given $\tau$, the full set of $\lambda$ solutions we are choosing from is rather rich; some even diverge or tend to zero (rather unphysical behaviours), or terminate at finite $r$. The more physical solutions tend, as $r \to 0$, to one of two $\lambda$ values: $1$ or $\Lambda$. This behaviour is unsurprising given that, for our current chosen $\alpha(r)$, any smooth $\lambda$ profile necessarily encodes a sharp tip in the activated surface unless $\lambda = 1$ or $\lambda = \Lambda$ at $r=0$~\cite{duffy2020defective}, and the target paraboloids have no such tips.

\begin{figure}[!h]
\centering
\includegraphics[width=\linewidth]{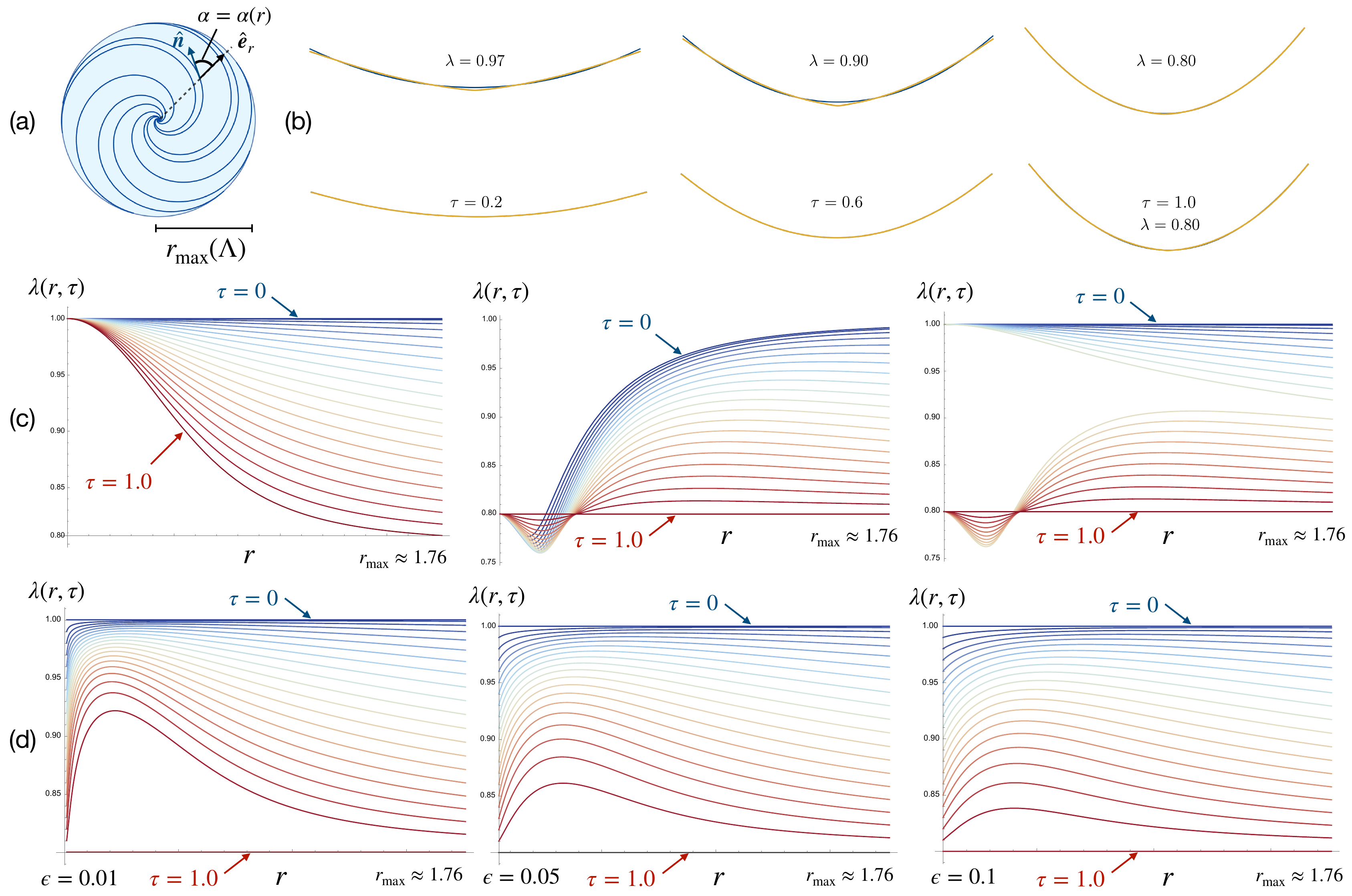}
  \caption{(a) Director field for a uniform-activation paraboloid, from (\ref{paraboloid director 1}) with $\Lambda = 0.8$, $\nu=1/2$, and $a = 1$. The domain extends to the maximal radius $r_{\max}(\Lambda)\approx 1.76$. (b) Top: Evolution of surface cross sections arising from uniform activation of the director (a), from $\lambda=1$ to $\lambda=\Lambda$, as computed by MorphoShell (yellow). These visibly deviate from even the best-fitting parabolas (blue), until the final $\lambda = \Lambda$ is reached. These deviations are partly \textit{qualitative}, since the activated surfaces have sharp tips. (b) Bottom: Cross sections (yellow) arising instead from designed \textit{inhomogeneous} activation of the same director. Agreement with the target family of paraboloids (blue) is excellent, hence the target curves are mostly obscured. (c) Various $\lambda(r,\tau)$ solutions to (\ref{DE flat to curved rev}), selected by specifying 
  $\lambda(0,\tau)=1$ (left), 
  $\lambda(r_{\max}(\Lambda)/4,\tau)=(1-\tau)(\Lambda+1)/2+\tau\Lambda$ (centre), and 
  a piecewise combination (right):
  $\lambda(0,\tau)=1$ for $\tau\leq 0.55$ and $\lambda(r_{\max}(\Lambda)/4,\tau)=(1-\tau)(\Lambda+1)/2+\tau\Lambda$ for $\tau>0.55$. All solutions are plotted from $r=0$ to $r_{\max}(\Lambda)$. (d) Alternative solutions on annuli $\epsilon \leq r \leq r_{\max}(\Lambda)$, selected by specifying $\lambda(\epsilon,\tau)=(1-\tau)+\tau\Lambda$ for $\epsilon = 0.01$ (left), $\epsilon = 0.05$ (centre), and $\epsilon = 0.1$ (right).}
   \label{fig:parabolic_path}
\end{figure}

In Fig.~\ref{fig:parabolic_path}b we show the surfaces resulting from uniform activations of our chosen     $\alpha(r)$ (Fig.~\ref{fig:parabolic_path}a), alongside surfaces generated by spatially varying activation profiles satisfying (\ref{DE flat to curved rev}); the latter match the target paraboloids essentially perfectly, while the former deviate significantly, and \textit{qualitatively} due to their sharp (potentially extremely undesirable) tips. To give a sense of the kinds of $\lambda$ solutions (\ref{DE flat to curved rev}) generates, each plot in Fig.~\ref{fig:parabolic_path}c shows a family of solutions, corresponding to the family of target paraboloids: In the first we specify $\lambda=1$ at $r=0$ for all $\tau$, resulting in profiles that are non-uniform apart from at $\tau=0$ (the flat state). In the second we specify that, at an intermediate $r$ value, $\lambda$ evolves linearly with $\tau$ to reach $\Lambda$ at $\tau=1$, resulting in profiles that are non-uniform apart from at $\tau=1$. In the third we combine the preceding two protocols piecewise to produce profiles that are uniform both in the initial state $\tau=0$ state and the `final' $\tau=1$ state, and exhibit relatively small spatial variations in $\lambda$ at the intermediate $\tau$ values, which is likely experimentally desirable. 

The discontinuous jump in $\lambda$ as a function of $\tau$ in the third solution in Fig.~\ref{fig:parabolic_path}c is potentially undesirable in practice.
To avoid this feature while retaining uniform $\lambda$ at $\tau=0$ and $\tau=1$, we instead look for solutions on annular domains.
Fig.~\ref{fig:parabolic_path}d shows solutions on annular domains $0<\epsilon\leq r \leq r_{\max}(\Lambda)$, specifying $\lambda(\epsilon,\tau)=(1-\tau)+\tau\Lambda$, for three small $\epsilon$ values. In each case $\lambda$ is again uniform at $\tau=0$ and $\tau=1$, but now evolves continuously with $\tau$ between these extremes.
Such annular solutions naturally cannot account for the material at $r<\epsilon$, which must either be physically removed, or assigned some other $\lambda$ profile; however, if $\epsilon$ is sufficiently small, the treatment of the $r<\epsilon$ material becomes essentially irrelevant to the overall shape, and need not be chosen painstakingly. Indeed MorphoShell simulations (Fig.~\ref{fig:parabolic_path}b) show that annular $\lambda(r,\tau)$ solutions with $\epsilon \sim \text{thickness}$ can work well even on a full disk of material. There, in fact, numerical $\lambda$ values did not even have to be assigned within the $r<\epsilon$ region, due to finite mesh resolution.

Note that in all of the above protocols, for a given $\tau$, how large a patch of the target paraboloid is attained depends on which $\lambda$ solution is chosen; this may be a crucial consideration in some applications. Indeed this consideration could lead to alternative protocols for selecting favourable solutions from the multitude that exist.

\subsection{Evolution from a curved surface using spatiotemporally programmed activation}

Modern fabrication techniques such as 3D printing allow an unactivated shape-programmable surface to be curved rather than planar~\cite{lopez20184d, kotikian20183d, ambulo2017four, 4dPrintingWare,4dPrintingSanchez, saed2019molecularly}, with the goal then being to morph this curved surface into a different curved surface upon activation. Here we briefly generalize the previous two sections to this curved$\to\,$curved setting, solving the director design problem for uniform-activation transformation from one surface of revolution to another, then adapting equation (\ref{DE flat to curved rev}) to again programme spatiotemporal activation and thereby transition between the same two surfaces via a prescribed family of surfaces.

Consider two surfaces of revolution $\mathcal{T}_i$ and $\mathcal{T}_f$, given by $z_i=z_i(\rho_i)$ and $z_f=z_f(\rho_f)$, respectively. We first seek to inscribe $\mathcal{T}_i$ with a circularly symmetric director field that upon activation results in a transition to $\mathcal{T}_f$. Let $(r,\theta)$ denote geodesic polar coordinates on $\mathcal{T}_i$ and $\alpha=\alpha(r)$ denote the angle that the director makes with the geodesic radial direction on $\mathcal{T}_i$. 
For simplicity, we assume that $r=0$ when $\rho_i=0$ so that the surfaces intersect the axis of revolution as in the case of paraboloids. 

By calculating an expression for the arc length on the target surface and matching it with the corresponding arc length determined by the activated metric as before, we find that the director $\alpha(r)$ for this inverse design problem is determined by
\begin{equation}  \label{inverse design curved to curved rev}
    \frac{\mathrm{d}}{\mathrm{d}r}\bar{a}_{\theta\theta}=\frac{2\rho_i(r)\lambda^{1-\nu}}{\sqrt{1+\left(z_{\tau}'\left(\sqrt{\bar{a}_{\theta\theta}}\right)\right)^2}},
\end{equation}
where the geodesic radius $r$ is related to $\rho_i$ by
   \begin{equation}  
       r(\rho_i)=\int_0^{\rho_i}\sqrt{1+(z_i'(\rho))^2}\; \mathrm{d}\rho,
   \end{equation}
$\rho_i=\rho_i(r)$ is the inverse of the function $r=r(\rho_i)$, and $\bar{a}_{\theta\theta}=\rho_i(r)^2[\lambda^{-2\nu}-(\lambda^{-2\nu}-\lambda^2)\sin^2(\alpha)]$.
Further details on this derivation are provided in Supplementary Materials, Sec.~S1b.

Having programmed a director $\alpha(r)$ for the transition from $\mathcal{T}_i$ to $\mathcal{T}_f$ under spatially uniform activation to $\lambda=\Lambda$, we can replicate the analysis of Subsection~\ref{DE flat to curved rev} for a prescribed family of intermediate surfaces of revolution $z_{\tau}=z_{\tau}(\rho_{\tau})$, arriving at the following equation which determines the spatiotemporally varying activation $\lambda=\lambda(r,\tau)$  required to track the geometry of the intermediate surfaces during transition from $\mathcal{T}_i$ to $\mathcal{T}_f$:
\begin{align} \label{DE curved to curved rev}
&\frac{\mathrm{d}\lambda(r,\tau)}{\mathrm{d}r}=
\frac{\lambda}{\rho_i\left(\nu c^2- \lambda^{2(1+\nu)}s^2\right)}\times \nonumber \\
& \left[\frac{-\lambda^{1+\nu}}{{\sqrt{1+\left(z_{\tau}'\left(\sqrt{\rho_i^2(\lambda^2 s^2
   +\lambda^{-2\nu}c^2)}\right)\right)^2}}}+ 
    \rho_i \left(\lambda^{2(1+\nu)}-1\right)\alpha'
   cs + 
    2 \rho_i'\left(\lambda^{2(1+\nu)}s^2 +c^2\right)\right],
\end{align}
where $s=\sin\alpha$ and $c=\cos\alpha$.

\begin{figure}[!h]
\centering
\includegraphics[width=\linewidth]{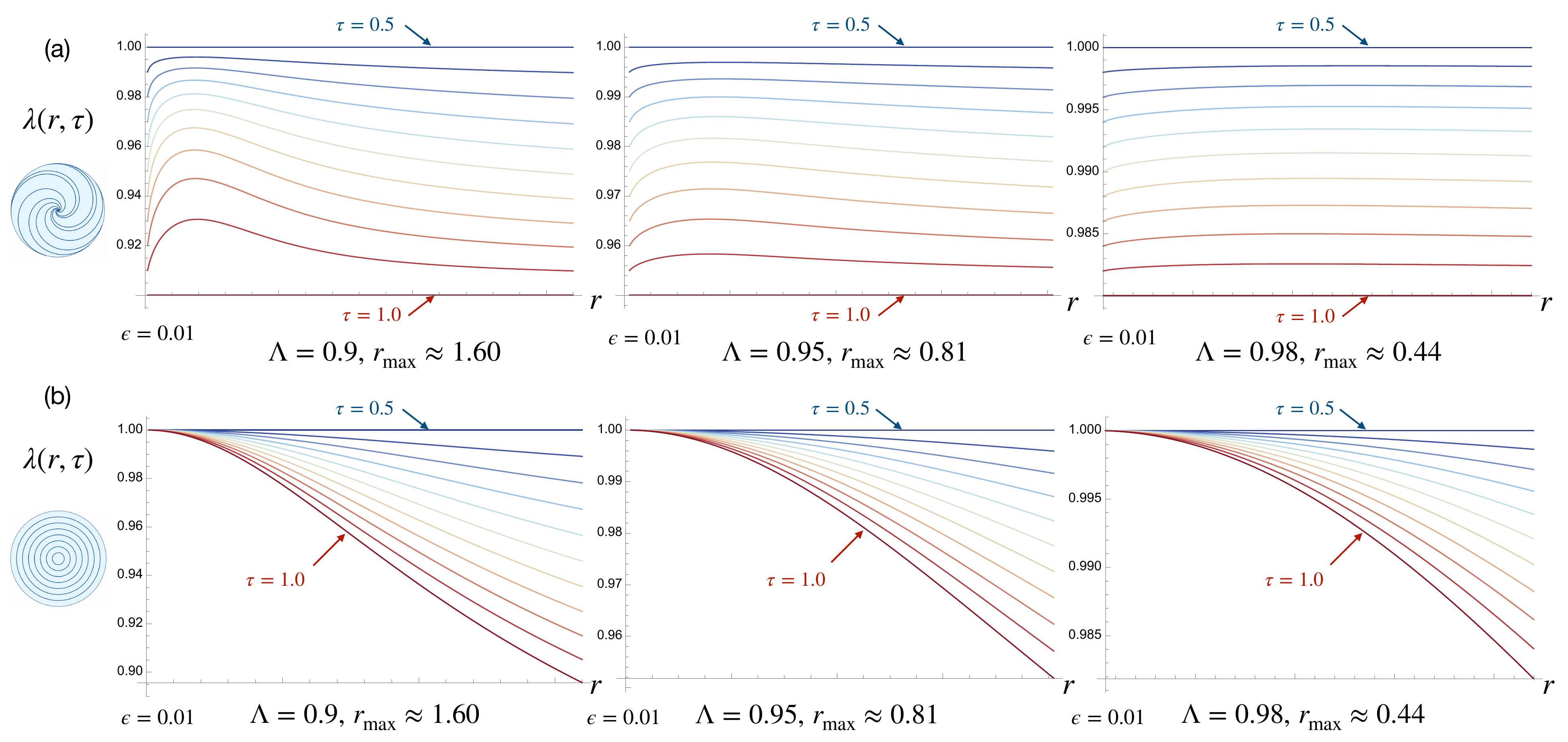}
  \caption{(a) Solutions $\lambda(r,\tau)$ of (\ref{DE curved to curved rev}) with $\nu=1/2$, to track the paraboloid family $z_{\tau}=\frac{1}{2}\tau\rho_{\tau}^2$ ($\tau\in[0.5,1]$). As $\Lambda$ becomes milder (left to right), the corresponding maximal radius $r_{\max}$ for the spiral director field (\ref{paraboloid_director_ctoc}) decreases, but the $\lambda$ solutions become increasingly homogeneous. We set $\lambda(\epsilon,\tau)$ to linearly interpolate from $\lambda=1$ to $\lambda=\Lambda$ for $\epsilon=0.01$, and the solutions are plotted for $\epsilon < r < r_{\max}$. (b) As in (a), except using an azimuthal director. These solutions exhibit much greater spatial inhomogeneity than (a) at the larger $\tau$ values. The $r_{\max}$ values in (b) are inherited from (a) merely to ease comparison; the azimuthal director can extend to infinity.}
   \label{fig:paraboloid_to_paraboloid}
\end{figure}

As an example, we consider a transition between two paraboloids given by $z_i=\frac{1}{2}a_i\rho_i^2$ and $z_f=\frac{1}{2}a_f\rho_f^2$, where our first task is to find a director profile that achieves the transformation under prescribed homogeneous stimulation. Solving (\ref{inverse design curved to curved rev}) for these surfaces yields the director $\alpha(r)$:
\begin{equation} \label{paraboloid_director_ctoc}
       \sin^2\alpha(r)= \frac{-\left( \Lambda^{1-\nu} (a_f/a_i)^2 
\left(\left(1+a_i^2\rho_i(r)^2\right)^{3/2}-1\right) + 1\right)^{2/3}+1+\Lambda^{-2\nu} a_f^2
   \rho_i(r)^2}{\left(\Lambda^{-2\nu}-\Lambda^2\right)a_f^2\rho_i(r)^2} ,
\end{equation}
where the target surface is attained with spatially uniform activation $\lambda=\Lambda$. We now seek to tune the spatial pattern of stimulation to track the desired intermediate surfaces; to do so, we substitute the above director field into (\ref{DE curved to curved rev}), which can then be solved to find spatiotemporally varying activation fields $\lambda(r,\tau)$ that encode target surfaces of revolution. See Fig.~\ref{fig:paraboloid_to_paraboloid}a for solutions $\lambda(r,\tau)$ that track the family of paraboloids $z_{\tau}=\frac{1}{2}\tau\rho_{\tau}^2$ for a transition from $a_i=0.5$ to $a_f=1.0$ for various choices of $\Lambda$. As in Fig.~\ref{fig:parabolic_path}d, we choose to set $\lambda(\epsilon,\tau)$ to linearly interpolate from $\lambda=1$ to $\lambda=\Lambda$ for a small value of $\epsilon>0$. 

Fig.~\ref{fig:paraboloid_to_paraboloid}b depicts the $\lambda(r,\tau)$ field necessary to track the same family of paraboloids via activation of an azimuthal director ($\alpha=\pi/2$) for comparison. Note that in contrast to the use of our spiral design, $\lambda$ must be maintained in a spatially inhomogeneous pattern at the final target surface\cite{giudici2022multiple}. 
Note also that in Fig.~\ref{fig:paraboloid_to_paraboloid}a, as the magnitude of the target deformation is reduced --- i.e.~$\Lambda$ is moved closer to 1 --- the spatiotemporal activation required to track the prescribed parabolic path with the spiral director design becomes increasingly spatially uniform for given $\tau$ (and for $r>\epsilon$). This contrasts with the solutions in Fig.~\ref{fig:paraboloid_to_paraboloid}b, whose azimuthal director field is less tailored to the parabolic target surface. On the other hand our spiral director can only be extended out to a finite $r_{\max}$, limiting the coverage that can be achieved, while the azimuthal director can be extended indefinitely and coupled with the appropriate spatiotemporal activation to achieve greater coverage of the target paraboloid, though this will nonetheless still be limited by practical restrictions on $\lambda$.

Thus the underlying choice of director field (material design) significantly affects the inhomogeneous $\lambda$ patterns that can encode a given target; a good designer may therefore wish to consider both in tandem, combining material design and spatiotemporal activation to find the most practical solutions for a given setup/application. In fact, in both the flat $\to$ curved and curved $\to$ curved cases, essentially \textit{any} axisymmetric director could be chosen for any given target; Fig.~\ref{fig:paraboloid_to_paraboloid} illustrates some of the trade-offs that can arise in this choice.

\section{Solving the general $\lambda$-inverse problem}\label{sec:general_inverse}

Sec.~\ref{sec:surfaces of revolution} showed how circularly symmetric inhomogeneous $\lambda$ fields can be designed to encode the geometries of surfaces of revolution, for a given circularly symmetric director pattern. As demonstrated there, prescribing \textit{paths} through shape space generally requires finding such spatiotemporally varying activation fields, and amounts to solving a metric inverse problem \textit{for each target surface}. Thus motivated, in this section we will solve this `$\lambda$-inverse' problem numerically in its full generality. 
However, we emphasise that while tracking a continuous family of targets provides strong motivation, one might well wish to solve this inverse problem for only a single target surface, or a discrete set of targets (e.g.~for reconfigurable haptics); there is \textit{no need} for targets to be drawn from any continuous family. Therefore we now omit any unnecessary mention of families from our statement of the general $\lambda$-inverse problem, which closely follows Sec.~\ref{sec: General Metric Inverse Design}:

\textbf{Problem III (Inverse design of $\lambda(x,y)$):}
\emph{Consider a flat sheet of material inscribed with coordinates~$x^\alpha$. Suppose this sheet will deform upon activation according to \eqref{eq:nematic metric}, with prescribed uniform $\nu$ and a prescribed director pattern $\hat{\bm{n}}(x^\alpha)$. Let there also be a target surface $\mathcal{T}$ inscribed with coordinates $X^\gamma$. Find $\lambda(x^\alpha)$ such that there exists a map $X^\gamma(x^\alpha)$ that yields $a_{\alpha\beta}=\bar{a}_{\alpha\beta}$ throughout some region of the flat sheet.}

\subsection{Formulation of the $\lambda$-inverse problem}

\begin{figure}[h]
\centering
\includegraphics[trim={0cm 0cm 0cm 0cm},clip, width=1.0\textwidth]{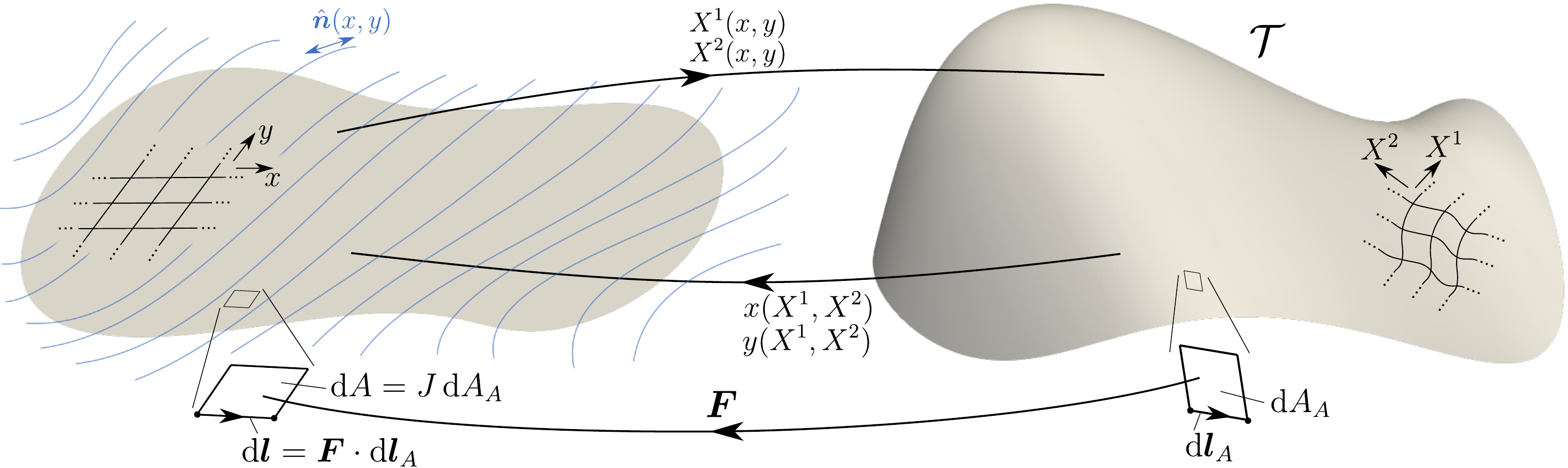}
\caption{Geometry of the $\lambda$-inverse problem. The target surface (right) is inscribed with coordinates $(X^1, X^2)$, and is mapped into the flat $x-y$ plane (left) by functions $x(X^1,X^2)$ and $y(X^1,X^2)$; the corresponding deformation gradient is $\bm{F}$. A fixed director field $\hat{\bm{n}}(x,y)$ (blue) lies in the $x-y$ plane. }
\label{fig:numerics_mappings}
\end{figure}

We are given a curved target surface $\mathcal{T}$ inscribed with coordinates $X^\gamma = (X^1, X^2)$, and a director field $\hat{\bm{n}} = (\cos (\psi(x,y)), \sin (\psi(x,y)))$ on the flat unactivated plane, for which we use Cartesian coordinates $x^\alpha = (x,y)$.
Let the matrix $\bm{G}$ contain the downstairs components of $\mathcal{T}$'s metric tensor, in the $(X^1, X^2)$ basis; i.e.~if the target surface has the position-vector form $\bm{R}(X^\gamma)$, then 
\begin{equation}
G_{\gamma \sigma} = \frac{\partial \bm{R}}{\partial X^\gamma} \cdot \frac{\partial \bm{R}}{\partial X^\sigma} . 
\label{eq:G_matrix}
\end{equation} 
Thus any infinitesimal squared length on $\mathcal{T}$ is given by $\mathrm{d}l_{\mathcal{T}}^2 = G_{\gamma \sigma} \, \mathrm{d}X^{\gamma} \mathrm{d}X^{\sigma}$, as in (\ref{eq:target_squared_length}).
We find it beneficial to work primarily with the 
target $\to$ planar deformation/mapping $(x(X^1, X^2)$, $y(X^1,X^2))$, with the corresponding deformation gradient (mixed components)
\begin{equation}
\bm{F} = 
\begin{pmatrix}
\partial x/\partial X^1 && \partial x/\partial X^2 \\
\partial y/\partial X^1 && \partial y/\partial X^2 
\end{pmatrix} ,
\label{eq:defgrad}
\end{equation}
and the areal scale factor $J = \sqrt{ \mathrm{det}(\bm{F} \bm{G}^{-1} \bm{F}^\mathrm{T})}$; see Fig.~\ref{fig:numerics_mappings}.
(Working directly with the deformation fields $x(X^1, X^2)$, $y(X^1,X^2)$ as unknowns guarantees that the compatibility requirement ${\mathrm{curl} \bm{F} = \bm{0}}$ is satisfied automatically.) We also define $\lambda(X^1, X^2)$ as a function on $\mathcal{T}$; naturally, this function can straightforwardly be converted into a function of $(x,y)$, given any invertible $(x(X^1, X^2), y(X^1,X^2))$.

Let $\bm{R}_\psi$ denote the standard $2 \times 2$ matrix that rotates vectors anticlockwise by $\psi$, and define $\bm{B} \equiv \mathrm{diag}(\lambda, \lambda^{-\nu})$. Then $\bm{R}_\psi \bm{B}^2 \bm{R}_\psi^\mathrm{T}$ is the matrix of downstairs components of $\bar{a}_{\alpha \beta}$ (\ref{eq:nematic metric}). Combining (\ref{eq:target_squared_length}), (\ref{eq:G_matrix}), and (\ref{eq:defgrad}) quickly reveals that $\bm{F}^{-\mathrm{T}} \bm{G} \bm{F}^\mathrm{-1}$ is the matrix of downstairs components $a_{\alpha \beta}$. Thus, solving Problem III amounts to solving $\bm{F}^{-\mathrm{T}} \bm{G} \bm{F}^\mathrm{-1} = \bm{R}_\psi \bm{B}^2 \bm{R}_\psi^\mathrm{T}$ or, equivalently, 
\begin{equation}
\bm{M} \equiv \bm{B} \, \bm{R}_\psi^\mathrm{T} \, \bm{F} \,  \bm{G}^{-1}  \bm{F}^\mathrm{T} \bm{R}_\psi \bm{B} = \bm{I}.
\label{eq:bulky_inverse_condition}
\end{equation}
We solve the above problem via an energy minimisation: One must simply construct a sensible energy (or `cost function') that penalises violations of the condition (\ref{eq:bulky_inverse_condition}). If a minimiser of that energy can then be found, \textit{and if it has zero energy}, then it is a solution of our inverse problem, since the condition (\ref{eq:bulky_inverse_condition}) is satisfied everywhere. The quantities to be minimised over are $\lambda(X^1, X^2)$ and the deformation fields $x(X^1, X^2)$, $y(X^1,X^2)$.

To select an appropriate energy, we note that any standard isotropic elastic energy-density function $W(\cdot )$ can be viewed as taking a symmetric positive semi-definite matrix (the `right Cauchy-Green tensor') as its argument, and penalising deviations of that argument from the identity matrix. 
Conveniently, any such energy density can be adapted for our purposes by simply taking $\bm{M}$ as the argument, because $\bm{M}$ is manifestly symmetric, and is in fact also positive semi-definite. To show the latter property, note that $\bm{M}$ is manifestly tensorial with respect to the $(X^1, X^2)$ coordinates. Thus, for ease, we can use locally Cartesian $(X^1, X^2)$ at any chosen point on the target, so that $\bm{G}^{-1} = \bm{I}$ there. Doing so, and defining $\bm{N} \equiv  \bm{B} \bm{R}_\psi^\mathrm{T} \bm{F}$, we have $\bm{M} = \bm{N} \bm{N}^\mathrm{T}$. Therefore $\bm{M}$ is positive semi-definite, since $\bm{v}^\mathrm{T} \bm{N} \bm{N}^\mathrm{T} \bm{v} = |\bm{N}^\mathrm{T} \bm{v}|^2 \geq 0$ for any vector $\bm{v}$.

We select the simple energy density function of a Neo-Hookean membrane, which (ignoring prefactors) can be written
\begin{equation}
W( \cdot) = \mathrm{tr}(\cdot)  + \mathrm{det}(\cdot)^{-1} -3.
\label{eq:neohookean_membrane}
\end{equation}
Take as the argument the positive semi-definite matrix $\bm{M}$ with eigenvalues $m_1$ and $m_2$, so $W(\bm{M}) = m_1+m_2 + 1/(m_1 m_2) - 3$. Since $m_1$ and $m_2$ are both non-negative, we see that indeed $W_s \geq 0$, with equality iff $m_1 = m_2 = 1$, i.e.~iff $\bm{M} = \bm{I}$. Thus (\ref{eq:neohookean_membrane}) indeed penalises violations of our inverse-problem condition (\ref{eq:bulky_inverse_condition}), as desired. Substituting our specific form of $\bm{M}$ yields the energy density that we use henceforth,
\begin{equation}
W_s = W(\bm{M}) = \mathrm{tr}\left( \bm{F} \,  \bm{G}^{-1}  \bm{F}^\mathrm{T} \bm{R}_\psi \, \bm{B}^2 \bm{R}_\psi^\mathrm{T} \right) + (J \lambda^{1-\nu})^{-2} - 3,
\label{eq:W_s}
\end{equation}
and integrating over area yields the energy.
To reiterate, if we can find fields $x(X^1, X^2)$, $y(X^1,X^2)$, $\lambda(X^1, X^2)$ such that $W_s=0$ over the entire target $\mathcal{T}$, then we have solved our inverse problem, and since $W_s \geq 0$ we can find solutions by minimising $\int W_s \, \mathrm{d}A$ over $x$, $y$, and $\lambda$.\\

Before moving onto our numerical minimisation procedure, we briefly discuss an alternative, equivalent formulation. The inverse problem corresponds exactly to a geometric condition on the target $\to$ planar mapping:
for every area element of $\mathcal{T}$, this mapping must first rigidly rotate and translate that element into the $x-y$ plane in some way, then stretch/compress by some factors $1/\lambda$ along $\hat{\bm{n}}$ and $\lambda^\nu$ orthogonally. At any point on $\mathcal{T}$ where we choose $(X^1,X^2)$ to be locally Cartesian, that condition reads $\bm{F} = \bm{R}_\psi \bm{B}^{-1} \bm{R}_\psi^\mathrm{T} \bm{Q}$, where $\bm{Q}$ is \textit{any} orthogonal matrix. We can absorb $\bm{R}_\psi^\mathrm{T}$ into $\bm{Q}$ without loss of generality, so the condition becomes $\bm{F} = \bm{R}_\psi \bm{B}^{-1} \bm{Q}$. 
(If non-Cartesian $(X^1,X^2)$ are used instead, then $\bm{Q}$ retains its geometric (rigid-body) interpretation, but can be any matrix satisfying $\bm{Q}^\mathrm{T} \bm{Q} = \bm{G}$.)
The above condition relating $\bm{F}$ and $\bm{B}$ leads to a formulation of the inverse problem as a system of two first-order nonlinear PDEs for two unknown fields (Supplementary Materials Sec.~S2). Thus the number of equations equals the number of unknowns, providing a strong expectation that the inverse problem does indeed have solutions. Furthermore, the PDEs facilitate a characterisation of the inverse problem as elliptic or hyperbolic for $\nu < 0$ and $\nu > 0$ respectively; a consequence is discussed in Sec.~\ref{subsec:bending_channel}. Notably, the character of PDEs describing planar kirigami deformations has been shown to depend on the sign of an effective Poisson ratio in exactly the same fashion~\cite{zheng2022continuum}.

\color{black}

\subsection{Numerical implementation}

Our formulation based on minimising (\ref{eq:W_s}) facilitates simple numerics:
We represent the target surface as a mesh of triangles, then flatten this mesh into the $x-y$ plane via some reasonable initial flattening. (In practice, it is more convenient to begin with a mesh in the $x-y$ plane, then generate the target mesh via a deformation of the planar mesh.) We take the $x$ and $y$ values at each node (i.e.~node positions in the plane) to be discrete degrees of freedom, linearly interpolating to define $x(X,Y)$, $y(X,Y)$ within each triangle, where $(X,Y)$ are Cartesian coordinates for the triangle's face in the target surface. We take $\lambda(X,Y)$ to be constant over each triangle, with its value being a discrete degree of freedom. (This discretization corresponds to $P1$ finite elements for $x$ and $y$, and $P0$ elements for $\lambda$.) Using a single quadrature point at each triangle's centroid, $\int W_s \, \mathrm{d}A$ becomes a sum over triangles, which we differentiate analytically with respect to the degrees of freedom to obtain `forces' on them. Thus, after assigning a mass to each, we evolve the degrees of freedom via lightly linearly damped Newtonian dynamics, using semi-implicit Euler time integration, terminating when the velocities drop below a small threshold. As mentioned earlier, it is then simple to convert $\lambda(X,Y)$ to a function of $(x,y)$, using the computed maps $x(X,Y)$, $y(X,Y)$.

There are typically infinitely many $W_s=0$ solutions for a given target. This non-uniqueness manifests most strikingly in the form of a `soft' (Nambu-Goldstone) mode that can be illustrated by considering a flat square target and a uniform horizontal director: the inverse problem can be solved by taking \textit{any} uniform $\lambda$ and a suitable rectangle with its long axis parallel to the director, since the $x-y$ domain is allowed to vary in the minimisation. Most members of such a solution family exhibit extreme $\lambda$ values, which are likely unattainable experimentally and also correspond to extreme mesh distortions in numerics. As in the conformal case~\cite{crane_bff}, there are many ways one can try to make the solution unique, or at least much less degenerate.
A simple one, directly addressing the problem of extreme $\lambda$ values, is to add a `tie-breaking' term $W_r$ to the energy density, penalising deviations of $\lambda$ from 1:
\begin{equation}
W_r = \Gamma \left( \lambda^2 + \frac{1}{\lambda^2} - 2 \right) ,
\label{eq:regularization_energy_density}
\end{equation}
where the prefactor $\Gamma$ is a positive tunable parameter. The total energy becomes $\int (W_s + W_r)\mathrm{d}A$, and keeping $\Gamma$ very small ensures that the minimiser will have $W_s = 0$ to an excellent approximation.\\

\section{Numerical $\lambda$-inverse examples}\label{sec:examples}

We now demonstrate our general numerical solver by applying it to two highly non-trivial examples. In each example a single target surface would suffice for demonstration, but we instead choose several from a family, in keeping with the paradigm of preceding sections, and with an eye towards real-world applications.

\color{black}

\subsection{A `twister'}

\begin{figure}[h]
	\centering
	\includegraphics[trim={0cm 0cm 0cm 0cm},clip, width=1.0\columnwidth]{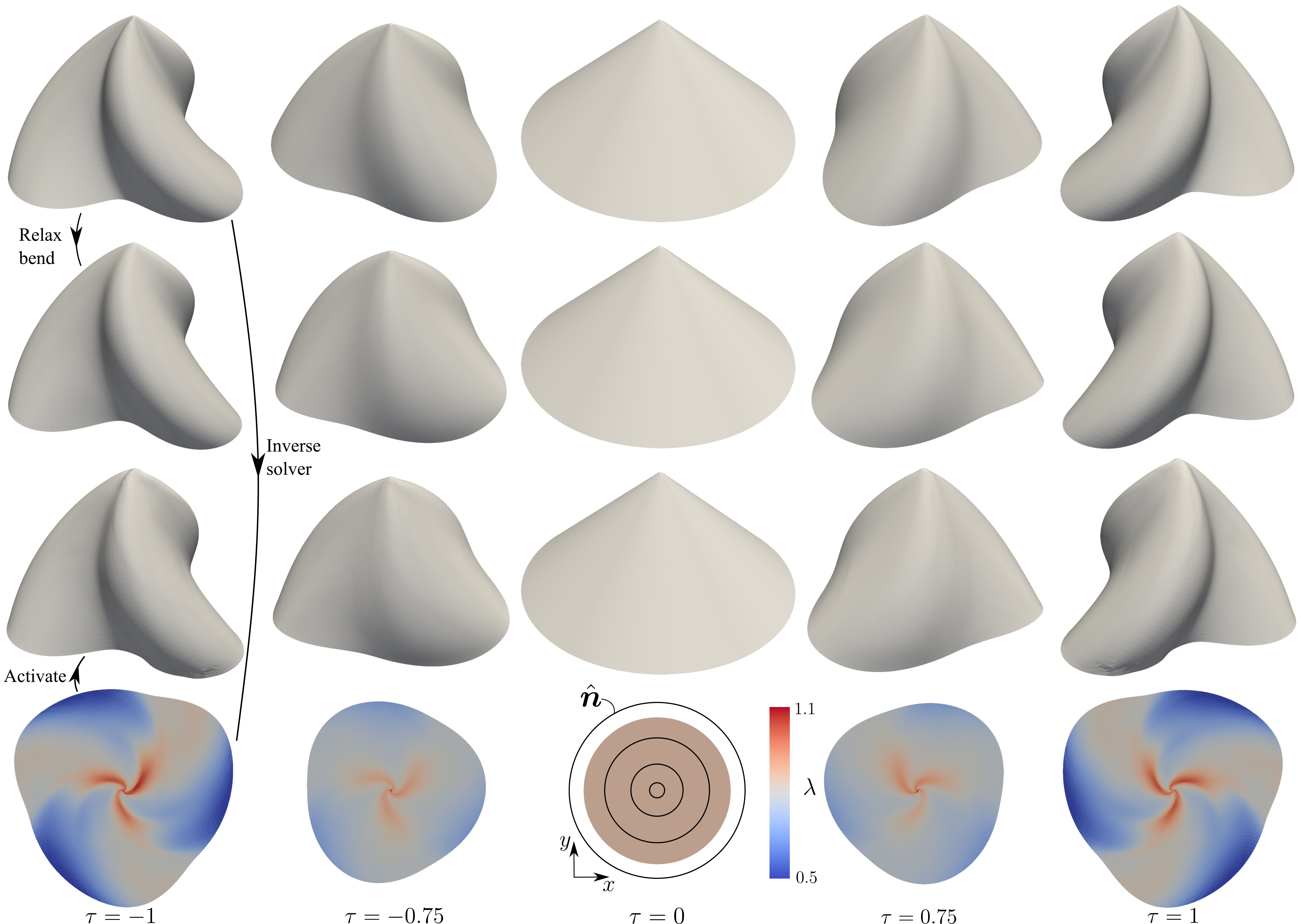}
	\caption{Row 1: Family of target `twister' surfaces parameterized by $\tau$, with $\tau=0$ corresponding to an LCE cone ($\nu=1/2$, $\Lambda = 0.85$) activated from a unit disk with azimuthal director $\hat{\bm{n}}$. Note that handedness switches with the sign of $\tau$. Row 2: MorphoShell bend-relaxations of the target surfaces, i.e.~approximately bend-minimising isometries (thickness $t=0.01$). Evidently the extent of bend relaxation is mild.  Row 3: MorphoShell activations of our inverse-problem solutions, exhibiting near-perfect visual agreement with row 2. Row 4: The numerical $\lambda(x,y)$ solutions that were activated to produce row 3. Note that $\hat{\bm{n}}$ must be defined in any area that the flattened sheet happens to explore during energy minimization; azimuthal $\hat{\bm{n}}$ naturally avoids any issue in this regard.}
	\label{fig:twister}
\end{figure}

We take $\hat{\bm{n}}$ to be azimuthal (a +1 topological defect), and construct a twisting family of target surfaces parameterized by $\tau \in [-1, 1]$, choosing the $\tau=0$ surface to be the cone of semi-angle $\beta = \arcsin(\Lambda^{1+\nu})$ produced by activating a unit disk activated with uniform $\lambda = 0.85 \equiv \Lambda $. We use 3D cylindrical coordinates $(\rho, z, \phi)$ for the embedding space, and inscribe the targets with coordinates $(X^1, X^2) = (R, \Theta)$, where $R \in [0, 1]$ and $\Theta \in [0, 2\pi)$. Then the family has the following form:
\begin{align}
\rho(R, \Theta) &= \rho_c(R, \Theta)  \left( 1 + 0.3 \, \tau^2  \sin\left( 3 \Theta \right) \right) , \notag \\
z(R, \Theta) &= z_c(R, \Theta)  \left( 1 + \tau^2 \,  R \right) , \notag \\
\phi(R, \Theta) &= \phi_c(R, \Theta) + \tau \pi z(R, \Theta) / 3 ,
\end{align}
where 
\begin{align}
\rho_c(R, \Theta) &= \Lambda R, \notag \\
z_c(R, \Theta) &= -\Lambda R / \tan \beta, \notag \\
\phi_c(R, \Theta) &= \Theta
\end{align}
describe the cone. Taking $\nu=1/2$ (as in LCEs), the $\tau = -1, -0.75,0, 0.75, 1$ surfaces are shown in Fig.~\ref{fig:twister} (row 1). Beyond their visual appeal, such surfaces could be used as active flow guides: A fluid deposited at the tip will flow down the flanks, predominantly following the curved valleys, and will therefore depart from the base in a clockwise sense for $\tau>0$ and an anticlockwise sense for $\tau<0$. This mechanism could be used to generate vortex-like flows in a shallow bath of fluid in which the shell's base is submerged, with the vorticity's sign controlled wirelessly via illumination. If a small hole were introduced at the tip, an active funnel with similar functionality would result.

Pure-bend deviations from target surfaces are always a potential pitfall that purely metric-based shape programming cannot directly address: as mentioned earlier, even if activation generates the \textit{metric} of the target surface correctly, the actual \textit{shape} adopted upon activation will be a bend-minimising isometry, which may be rather different. Crucially, by computing the bend-minimising isometries of our target surfaces using MorphoShell, we confirm that this is \textit{not} an issue in this case, as the bend-minimizing isometries barely differ from the targets (Fig.~\ref{fig:twister} row 2). (In contrast, a `twister' with six lobes rather than three suffers a severe symmetry-breaking bend relaxation that would likely affect functionality; see Supplementary Materials Fig.~S2.)

Setting $\Gamma = 0.005$ in (\ref{eq:regularization_energy_density}),
our solver yields good inverse solutions (Fig.~\ref{fig:twister} row 4) despite using somewhat coarse meshes: in every case $W_s < 0.01$ in at least 99.9\% of the $\sim 18000$ mesh triangles. `Heat maps' of $W_s$ are given in Supplementary Materials Fig.~S3. Activating our solutions with MorphoShell (Fig.~\ref{fig:twister} row 3), we find that the resulting bend minimisers agree extremely well with the bend-minimising isometries of the target surfaces; they are almost indistinguishable visually. Note that neither the central discontinuity in $\hat{\bm{n}}$, nor the sharp tips of the target surfaces, presented any barriers to our success.

Having calculated the inverse solution for $\tau=1$, say, an alternative appealing family of surfaces and a route for its traversal present themselves without further effort: by simply rotating the pattern of activation $\lambda$ with respect to the physical flat domain we can traverse a family of target surfaces that are rotations about $z$ of the top-right surface in Fig.~\ref{fig:twister}. Such a path through shape space is \textit{non-reciprocal}, so could be used for swimming through a viscous fluid~\cite{purcell2014life} or driving its flow. Note, however, that although the shell's motion might visually appear to be rotational, this appearance \textit{might} be somewhat deceiving depending on the fluid-structure interaction, and care must be taken. The case of activation in free space without fluid illustrates the subtlety: even in that case the motion would \textit{resemble} a rigid-body rotation, but certainly could not \textit{be} one, due to conservation of angular momentum.

The $\lambda(x,y)$ patterns in Fig.~\ref{fig:twister} row 4 successfully solve our inverse problem, but they have a potentially unfortunate feature: the $x-y$ domain is different for each solution in the family, so strictly speaking a single physical sample can only be activated according to one of the solutions. To circumvent this irritating issue, one could activate only subregions of the inverse solutions using a single `cookie cutter', the simplest being the largest disk common to all. The material outside this common domain would ideally be physically cut away and discarded. Naturally this approach will result in activated samples that achieve only partial coverage of the target surfaces.

\subsection{An active channel}
\label{subsec:bending_channel}

A more direct way to control the $x-y$ domain of our inverse-problem solutions is to specify (`pin') the shape of the boundary as a constraint (i.e.~impose a relation between the boundary values of $x$ and $y$). This strategy is attractive experimentally, because it allows a \textit{single} experimental patterned sample to attain the metric of essentially \textit{any} target surface, via an appropriate pattern of activation yielded by our numerics; thus a single sample can be used for arbitrarily many targets. 

\begin{figure}[h]
	\centering
	\includegraphics[trim={0cm 0cm 0cm 0cm},clip, width=1.0\columnwidth]{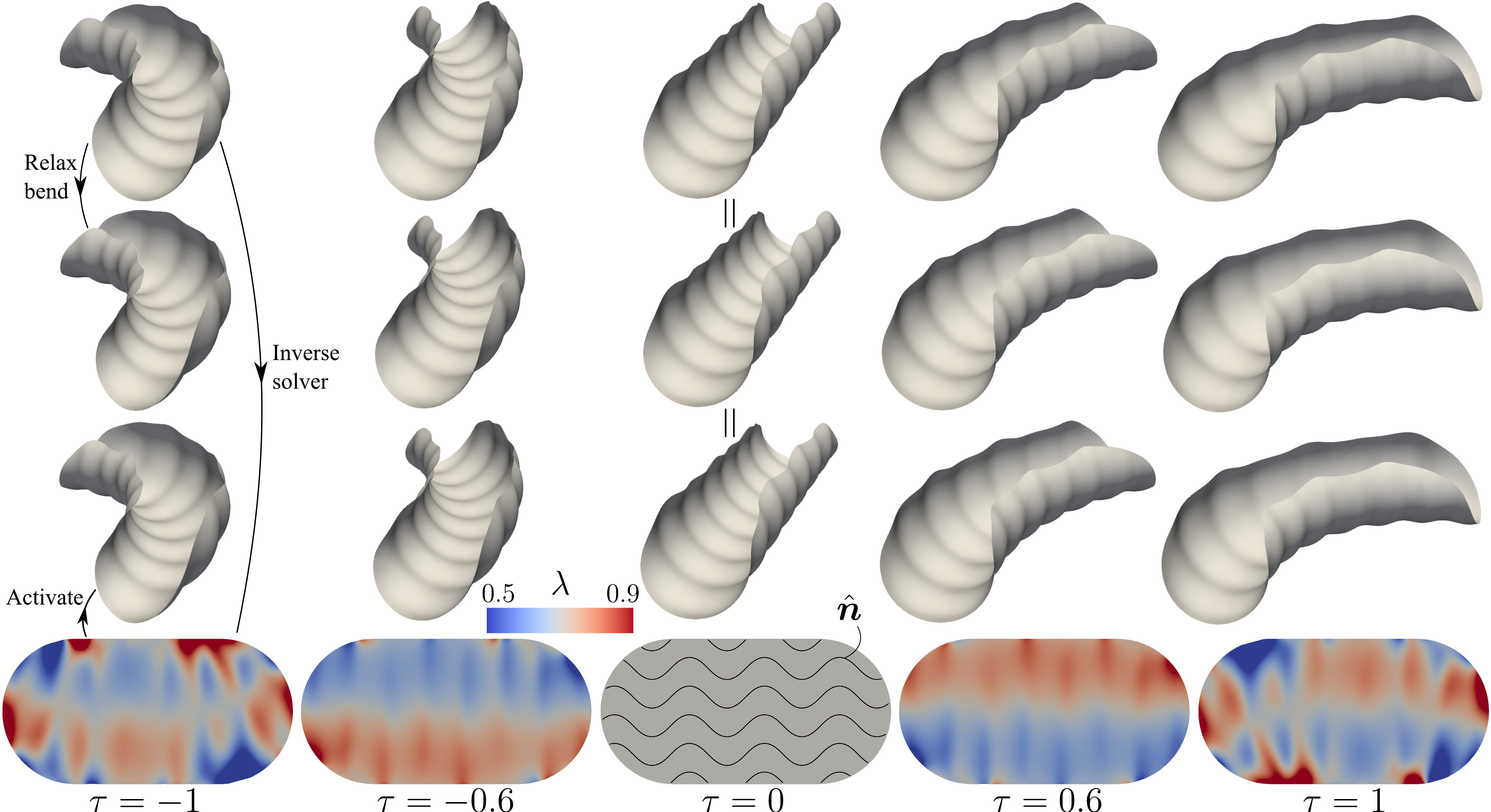}
	\caption{Row 1: Family of target channel surfaces parameterized by $\tau$, with $\tau=0$ corresponding to $\nu=-1/2$, $\Lambda = 0.7$ activation of the director pattern $\psi = \pi \cos(6x) /4 $ on a pill-shaped domain of height 2 and total length 4. For $\tau \neq0$ the major radius $\rho_m = 1.5/\tau$. Row 2: MorphoShell bend-relaxations of the target surfaces (thickness $t=0.01$). Row 3: MorphoShell activations of our $\lambda(x,y)$ inverse solutions, agreeing nearly perfectly with row 2. Row 4: The numerical inverse solutions, calculated while restricting the domain shape to be the same pill as in the original uniform-$\lambda$ design. The most extreme $\lambda$ values were $\sim 0.05$ and $\sim 2$, occurring in small regions of the $|\tau|=1$ solutions. Thus some colours above clip/saturate the scale; for versions without clipping see Supplementary Materials Fig.~S4.}
	\label{fig:bending_channel}
\end{figure}

With an eye towards fluidic applications once again, we now present another example, demonstrating both boundary pinning and a semi-qualitative inverse-design workflow. This workflow is enabled by MorphoShell, which allows one to rapidly calculate approximate activations of director patterns, and thereby prototype designs easily. Suppose we seek to direct/channel a flowing fluid left or right using an active shell. To do so, we can first qualitatively target something approximating a sector of a cylinder as the `central' $\tau=0$ member of a family, to be attained upon uniform activation of some director pattern. We can then bend this channel one way or the other to generate a family of target surfaces (Fig.~\ref{fig:bending_channel} row 1), whose metrics will be achieved via patterned activation $\lambda(x,y)$ that we calculate with our numerical solver.

An appropriate $\tau=0$ surface is not immediately obvious: a sector of a cylinder is Gauss flat, so will always bend-relax to the plane, and thus cannot be attained by merely activating a programmed metric (introducing a preferred curvature would be necessary). However, a `ribbed' channel circumvents this problem, since it bears Gauss curvature, and therefore cannot be flattened without stretch. As a by-product, its activation will be much more robust than that of a Gauss-flat channel with preferred curvature, because stretch is generally far more energetically costly than bend in thin shells. The simplest conceivable ribbed channel is a partial surface of revolution with a periodic oscillatory profile in the axial direction. These symmetries immediately suggest that we should trial a director pattern that is translationally invariant in one direction, while oscillating in the other, such as $\psi = b \cos( c x) $. We conduct trial-and-error MorphoShell activations of such patterns for a reasonable uniform $\Lambda = 0.7$, setting $\nu=-1/2$. 
These simulations quickly reveal sensible choices of $b$, $c$, and domain, finally yielding the $\tau =0 $ channel in Fig.~\ref{fig:bending_channel} (row 1), whose axis is parallel to the $x$-axis. Note that domains of smaller $y$-extent lead to shallower trough-like channels, while domains of greater $y$-extent allow one to wrap the activated surface's cylindrical envelope azimuthally as many times as desired (the result rather resembling an elephant's trunk).

Denoting the 3D Cartesian coordinates of points on the $\tau = 0 $ channel by $(\tilde{x},\, \tilde{y}, \, \tilde{z})$, it is simple to generate a family of bent channel target surfaces, by assigning each point 3D cylindrical coordinates
\begin{align}
\rho &= \rho_m - \tilde{y} , \\
\phi &= \tilde{x}/\rho_m -\pi/2, \\
z &= \tilde{z},
\end{align}
taking the `major radius' $\rho_m \propto 1/\tau$. As before, MorphoShell reveals that these targets are indeed appropriate for metric programming, in that they deviate only mildly from their bend-minimising (mechanically favoured) isometries (Fig.~\ref{fig:bending_channel} row 2).

Running our inverse solver, we pin the $x-y$ domain boundary to match that of the original ($\tau=0$) uniform-$\lambda$ design; a `pill' shape in this case. To do so, we apply an extra force to each node on the mesh boundary, whose size is proportional to the (shortest) distance from the node to the desired boundary curve, and which points from the node towards the closest point on that curve. As before, we find good solutions: in each case $W_s < 0.001$ in at least 99.8\% of mesh triangles (`heat maps' of $W_s$ are provided in Supplementary Materials Fig.~S4). Furthermore, activating these solutions with MorphoShell (Fig.~\ref{fig:bending_channel} row 3), we find again that the resulting surfaces are visually indistinguishable from the targets' bend relaxations. Some of our solutions do exhibit rather extreme $\lambda$ values in small regions (e.g.\ $\lambda \sim 0.05$); if necessary this could be ameliorated by increasing $\Gamma$, at the cost of solving the inverse problem less exactly. 
\\

Numerical boundary pinning was thus highly successful in the above design, allowing us to morph a single physical sample into various (entire) target surfaces. However, further numerical explorations reveal some limitations:
First, the boundary constraint typically appears to prevent the existence of inverse solutions ($W_s=0$) if $\nu > 0$, which likely reflects the fact that the underlying PDE system is hyperbolic in that case, while it is elliptic for $\nu < 0$, as we show in Supplementary Materials Sec.~S2. Second, if the target has multiple boundaries (non-disk topology, as in~\cite{duffy2021metric}), only one of them can be subjected to the shape constraint, otherwise an inverse solution will typically not exist. Third, if the director pattern contains a topological defect
discontinuity (as in our earlier `twister' example), pinning the boundary often seems to yield a $\lambda$ field that diverges at the defect. (The divergence is of course `cut-off'/regularized in numerics by finite mesh resolution and/or $W_r$ (\ref{eq:regularization_energy_density}).) We attribute this issue to the fact that, if a smooth $\lambda$ field is to take a finite positive value at a director defect, only a few discrete values encode the correct singular Gauss curvature there ($=0$ for a smooth target)~\cite{duffy2020defective}, effectively placing extra constraints on smooth $\lambda$ solutions that may not be compatible with a boundary constraint. Other director discontinuities such as `seams' will likely induce similar issues, since they also generically encode concentrated Gauss curvature~\cite{feng2022interfacial}. We suspect further pinning issues will arise when boundaries have sharp intrinsic corners either in the target or in their flat pinned state. If desired, one can proceed with boundary pinning while ignoring all of the above limitations; the resulting energy minimum \textit{may} be close enough to zero for the minimiser to constitute a `good enough' solution to the inverse problem in some cases.

Some intuition for the second and third limitations above emerges if we analytically target a full or truncated cone of semi-angle $\beta$ and boundary radius $\rho_b$, via azimuthal director and axisymmetric $\lambda(r)$: Considering the required deformation of an infinitesimal annular element leads to $r \lambda' + \lambda = \lambda^{-\nu} \sin \beta $, with particular solutions $\lambda^{1+\nu} = \sin \beta$ and $\lambda^{1+\nu} = \sin \beta + c / r^{1+\nu}$ for an arbitrary constant $c$. One solution or the other must be taken, since they manifestly cannot be continuously glued together piecewise. Given $\rho_b$, the first solution offers no freedom to choose the flat-state outer (boundary) radius $r_o$, since $\rho=\lambda r$. The second solution does allow choosing $r_o$ by fixing $c$, but has $\lambda$ diverging as $r \to 0$; or in the case of an annular domain (for a truncated-cone target) it offers no further freedom to choose the inner $r_i$.
\\

\subsection{Discussion}

In solving any metric inverse problem, (rarely discussed) issues of domain/coverage inevitably raise their heads: One would ideally like to cover the entire target surface while also controlling the shape of the inverse solution's domain (i.e.~its boundary curve). We have shown that in the $\lambda$-inverse problem this is possible only in some cases, which we have begun to map out. Beyond such cases, less freedom is available for solutions, so sacrifices must be made. The seriousness of this limitation is undoubtedly context dependent. A comprehensive exploration of such issues would be interesting and worthwhile, in both the $\lambda$-inverse problem and the $\hat{\bm{n}}$-inverse problem~\cite{aharoni2018universal, griniasty2019curved, pauly_2021_inverse, itay_multivalued, aharoni2023plce}. Moreover, patterning both $\lambda$ and $\nu$ independently (if feasible) would offer larger, less restricted solution spaces; 
our numerical inverse solver could handle such systems straightforwardly by promoting $\nu$ to a DOF $\nu(X^1, X^2)$.

The issue of bend has also often been glossed in previous works on metric-based shape programming, but is important and deserves discussion: In general, target surfaces have many isometries. Which of these is actually mechanically selected upon activation is governed by bend energy, which penalises deviations in the surface's curvature tensor $b$ (`second fundamental form') from some preferred/`programmed' $\bar{b}$. Thus one sensible strategy, if feasible, is to programme $\bar{b}$ in addition to the metric $\bar{a}$, to ensure that the desired isometry is adopted. This can be achieved by programming spontaneous deformations that vary through the sheet thickness~\cite{aharoni2018universal, mahaElasticBilayers, barnes2019direct, plucinsky2018patterning}, or orienting the direction of minimum bending stiffness in anisotropic materials~\cite{pauly_2021_inverse}. We have instead programmed only $\bar{a}$ while keeping $\bar{b}=0$, but we have judiciously chosen target surfaces that will nonetheless be energetically selected (i.e.~are bend-minimising) to a good approximation. While this paradigm is undoubtedly more limited, it is far simpler, and does not preclude successful programming of striking and functional surfaces, as our examples demonstrate. Conveniently, as well as reducing metric programming problems to ODEs, surfaces of revolution are their own bend-minimising isometries, so our target surfaces in Sec.~\ref{sec: paraboloid problem} and Sec.~\ref{sec:surfaces of revolution} were automatically appropriate. In Sec.~\ref{sec:examples} we moved on to consider more general surfaces (Figs~\ref{fig:twister} and \ref{fig:bending_channel}), and our approach was assisted by MorphoShell, with which it is easy to compute bend-minimising isometries of potential targets in trial-and-error fashion. (For example, these computations led us to choose a `twister' with three rather than six lobes; compare Figs~\ref{fig:twister} and Supplementary Fig.~S2.) Our target choices were also guided by an expectation that surfaces with more pronounced features and more Gauss curvature are typically likely to undergo less severe bend-relaxing deformations. In practice, even significant bend-relaxing deformations might not affect the \textit{functionalities} of some active surfaces; these would thus constitute an even larger class of targets amenable to the $\bar{b}=0$ paradigm.

We make four closing comments on the applicability of our framework going forward: First, while our examples in Figs~\ref{fig:twister} and \ref{fig:bending_channel} were not completely free of symmetry, this was merely to retain some simplicity and facilitate intuition; our numerics are fully general and do not utilize symmetries in any way. Second, our inverse solutions typically have regions where $\lambda > 1$, which may be undesirable or inaccessible for some experimental systems (e.g.~LCEs). This issue can be addressed straightforwardly by (as applicable) replacing (\ref{eq:regularization_energy_density}) with a form whose minimum lies at some $\lambda$ sufficiently far below 1, or scaling a pinned boundary shape in overall size. Third, in practice the patterned stimulus underlying a spatially varying $\lambda$ might be applied in an `Eulerian' manner, meaning that an experimenter must know how to maintain/adjust $\lambda$ as a function of ambient 3D spatial position. Patterned illumination from a projector exemplifies this situation. In such cases, $\lambda(x,y)$ patterns over the unactivated $x-y$ domain (as shown in Figs~\ref{fig:twister} and \ref{fig:bending_channel}) are insufficient; as the sample deforms, the experimenter must also keep track of where each material point $(x,y)$ of the unactivated sample is currently located in 3D space. They could do so by imaging visible marks on the sample, or by simulating the activation/deformation process alongside their experiment (using MorphoShell, for example). Fourth, our approach could also be applied to many other metric inverse problems. Indeed (\ref{eq:W_s}) is closely related to the LCE trace formula~\cite{warnerbook}, and our formulation is broadly inspired by the beautiful experimental strategy of direct mechanical programming~\cite{barnes2019direct}, in which an LCE sample is physically pressed onto a target shape, while allowing the material to adjust (minimise energy) subject to that shape constraint. In such LCE experiments $\lambda$ cannot adjust, while $\hat{\bm{n}}$ can (via `soft elasticity'~\cite{warnerbook}). In that spirit, we could just as well tackle the `$\hat{\bm{n}}$-inverse problem' by fixing $\lambda$ while promoting the director angle $\psi$ to a DOF $\psi(X^1, X^2)$ --- an alternative to the formulations in refs~\cite{aharoni2018universal, griniasty2019curved, pauly_2021_inverse, aharoni2023plce}. Or we could consider materials with more than one DOF in $\bar{a}$, accordingly minimising our energy over more fields, e.g.~$x$, $y$, $\lambda$, and $\nu$.

\section{Conclusions}

One successful paradigm for shape programming of shells is nematic `metric mechanics': patterning the principal axis $\hat{\bm{n}}$ of an in-plane spontaneous deformation to encode the metric of a target surface, e.g. using LCE/Gs~\cite{warner2020topographic, warner_disclination, aharoni2014geometry, de2012engineering, ware2015voxelated, ambulo2017four, 4dPrintingWare, 4dPrintingSanchez, duffy2021metric, duffy2021shape} or  pneumatics~\cite{siefert2019bio, siefert2020programming, siefert2020inflationary}. Usually the deformation magnitudes ($\lambda$ and $\lambda^{-\nu}$) are spatially uniform, and in one-to-one correspondence with some external stimulus such as temperature or pressure; then the target is achieved at some particular value of the stimulus. We have shown that at other values of the stimulus, such systems generally adopt shapes that deviate from the target in \textit{qualitative} ways that may be undesirable, e.g.~their shapes may not be in the same (desired) geometric family as the target, may not offer the same functionalities, and may even have a lower degree of smoothness. Thus, if one wishes to track a family of target surfaces, greater control over the spontaneous deformation is required. Unfortunately, but intuitively, merely allowing the deformation magnitudes to be spatially inhomogeneous is not sufficient to solve the problem, if they are still controlled by a time-varying global stimulus parameter via a time-invariant relationship; achieving multiple targets becomes possible, but not infinitely many~\cite{itay_multivalued}.

Instead, full spatiotemporal control of at least one parameter per material point is required, i.e.~one must solve a family of metric inverse design problems (one per target).
The parameter $\hat{\bm{n}}$ is usually not mutable `on the fly', so is not a promising candidate for spatiotemporal control, but current experimental systems do offer the possibility of making patterned adjustments of a single deformation-magnitude parameter $\lambda$ in real time, e.g.~via patterned illumination~\cite{kuenstler2020reconfiguring, giudici2022multiple}. We have therefore worked in that paradigm, solving the $\lambda$-inverse problem semi-analytically for some special cases, and also in general via a novel numerical scheme. We emphasise that one might well wish to solve this inverse problem for only a single target surface, or a discrete set of targets (e.g.~for reconfigurable haptics); tracking a \textit{family} of targets was our motivation, but is merely one (rather advanced) application of the scheme.

This application is, however, perhaps the most exciting, in that it allows one to programme \textit{paths} through shape space (modulo isometric deformations). Naturally, the ability to traverse such paths is extremely important for making versatile soft robots, e.g.~when operating in confined environments or performing complex tasks. Shape-morphing structures have been used as parts of soft robots before~\cite{mirvakili2018artificial, shah2021shape, yang2023morphing}, but varied complex motions are typically hard to achieve, requiring many shape-morphing components to be combined, often in ways that are not easily systematized. In contrast, the novelty of our spatiotemporally activated continuum systems is that a single component can accomplish essentially any motion, rivalling the richness of biological shape morphing.

Particularly notable is that closed non-reciprocal loops through shape space become straightforward to achieve, again enabling greatly expanded functionality in soft robotics, as well as the driving of viscous fluids, including swimming~\cite{purcell2014life}. In fact, \textit{almost all} loops are non-reciprocal, so constructing them is by no means challenging, though the enticing question of which loops are \textit{optimal} certainly is; the answer will certainly be application dependent. \\

If full spatiotemporal control of $\lambda$ is feasible, the specifics of the director field $\hat{\bm{n}}$ may seem unimportant; indeed roughly speaking, given $\hat{\bm{n}}$, any target can be achieved by activating a suitable domain with a suitable patterned $\lambda$. However, choosing $\hat{\bm{n}}$ judiciously might still be worthwhile; in other words a \textit{combination} of material design and spatiotemporal activation may be the best option. Suppose one designs $\hat{\bm{n}}$ to achieve a single target at a single uniform $\lambda$. First, a \textit{single} fabricated sample can then be used in two different `modes': patterned $\lambda$ can be used to track a family exactly if required, or simple uniform $\lambda$ can be used instead, offering exact matching of one target and approximate family tracking that may be accurate enough for some applications. Second, compared to less careful choices of $\hat{\bm{n}}$, the $\lambda$ field(s) required to achieve the desired target(s) will often be significantly less extreme or closer to homogeneous, which could be essential for experimental feasibility; see Fig.~\ref{fig:paraboloid_to_paraboloid} and the discussion thereof. 
In general, selecting a material design / activation pattern pair that is in some way optimal for a given setup may call for computational tools in the vein of~\cite{Spielberg2023}.\\

\enlargethispage{20pt}

\ack{C.M.~and D.D.\ acknowledge financial support from the School of Physical and Mathematical Sciences and the Talent Recruitment and Career Support (TRACS) Office at Nanyang Technological University (NTU). I.G.\ was supported by the Eric and Wendy Schmidt \textit{AI in Science Postdoctoral Fellowship}. J.~S.~B.\ was supported by a UKRI \textit{Future Leaders Fellowship} grant (Grant No.~MR/S017186/1).}

\vskip2pc

\bibliographystyle{RS}
\bibliography{references}

\end{document}